\documentclass[11pt]{article}
\usepackage{amssymb,cite,epsf,graphics,graphicx}
\begin{document}

\title{From S-matrices \\
to the Thermodynamic Bethe Ansatz}
\author{\textbf{A. Babichenko}\thanks{%
e-mail: babichen@wicc.weizmann.ac.il} \\
\\
\textit{Department of Particle Physics,}\\
\textit{\ Weizmann Institute of Science, Rehovot 76100, Israel}.}
\date{November 2004}
\maketitle

\begin{abstract}
We derive the TBA system of equations from the S-matrix describing
integrable massive perturbation of the coset $G_{l}\times G_{m}/G_{l+m}$ by
the field $(1,1,adj)$ for all the infinite series of simple Lie algebras $%
G=A_{n},B_{n},C_{n},D_{n}$. In the cases $A_{n},C_{n}$, where the full
S-matrices are known, the derivation is exact, while the $B_{n},D_{n}$ cases
dictate some natural assumption about the form of the crossing-unitarizing
prefactor for any two fundamental representations of the algebras. In all
the cases the derived systems are transformed to the corresponding
functional $Y$-systems and shown to have the correct high temperature (UV)
asymptotic in the ground state, reproducing the correct central charge of
the coset. Some specific particular cases of the considered S-matrices are
discussed.
\end{abstract}

\section{Introduction}

Thermodynamic Bethe Ansatz is know to be one of the most impressive
achievements of two dimensional physics relating (in the high temperature
limit) the data of a perturbed conformal field theory (CFT) and an exact
factorizable S-matrix for relativistic integrable model with its spectrum
and analytical structure. Unfortunately not only S-matrix often has a
conjectural status, but also the TBA system corresponding to the S-matrix
(e.g. \cite{R}\cite{QRT}\cite{T}\cite{DTT}\cite{DPT}). The main obstacle in
exact derivation of TBA system from S-matrix is usually related to the
problem of diagonalization of transfer matrix (TM), which is especially a
non trivial issue, when S-matrix has internal degrees of freedom, i.e. is
non diagonal. In this paper we show how one can solve this technical problem
for relativistic integrable models using some facts, established in
investigation of trigonometric TMs of lattice models with Lie algebraic
symmetries. These results were obtained in the framework of algebraic Bethe
ansats \cite{BR}\cite{K}, TM functional relations and analytical Bethe
Ansatz \cite{KNS}. 

The method to incorporate lattice model as a ''carrier''
of magnonic degrees of freedom responsible for entire symmetries of a
relativistic integrable model, is an alternative to another method 
of explicit lattice light cone regularization of whole relativistic
model (see, e.g. \cite{DV}\cite{RS}). As far as we know, the
first time this program was successfully brought about by Hollowood in \cite
{H}, where he derived TBA system in the $A_{n}$ case. Another successfull 
implementation of this procedure was done in \cite{F} and \cite{BT}. 
In this paper we use this method applying it to the integrable models with
other Lie algebraic symmetries.

It is known for a long time \cite{ABL} that integrable quantum field
theories arising as perturbations of the coset CFT $\frac{G_{l}\times G_{m}}{%
G_{l+m}}$ by the operator $(1,1,adj)$ for different Lie algebras $G $ is a
wide class of models, with with $U_{q}(G^{(1)}) $ symmetric trigonometric
S-matrices , $q$ a root of unity. More precisely, it was conjectured in \cite
{ABL} that the massive S-matrix has the symmetry $U_{q_{1}}(G^{(1)})\otimes
U_{q_{2}}(G^{(1)})$, where $q_{1}=-e^{-i\pi /(l+g)},q_{1}=-e^{-i\pi /(m+g)}$%
, where $g$ is the dual Coxeter number of $G$. In other words, the S-matrix
is, up to a scalar factor, a tensor product of two trigonometric $U_{q}$
invariant integrable models at $q$ a root of unity, i.e. RSOS models. This
class of S-matrices contains some other integrable quantum field theories
(IQFT) with Yangian symmetry (rational S-matrices), as a subclasses with
specific choice of parameters $l,m$: Principal Chiral models (PCM): $%
l,m\rightarrow \infty $, Gross-Neveu models (GN)\footnote{%
Note that not any Lie algebraic symmetry may be a symmetry of GN model: its
impossible to construct \ a GN-like interaction of Majorana fermions with $%
Sp $ group.} and current-current perturbations of WZW models: $%
m=1,l\rightarrow \infty $. There is a lot of literature devoted to
investigation of these models and we won't cite it here. In the context of
the general $U_{q_{1}}(G^{(1)})\otimes U_{q_{2}}(G^{(1)})$ symmetry these
limits for $m,l$ mean some essential simplification in the S-matrix
structure, related to peculiarity of RSOS models and some identities
existing between them for low level of restriction, which we discuss in the
last section.

As we said, the main goal of this work is direct derivation of \ TBA
equations from the full S-matrix and their high temperature analysis. In
principle, such derivation is impossible, in $B_{n}$ and $D_{n}$ cases,
since the spectral decomposition of the S-matrices is unknown for arbitrary
two fundamental highest weights of the algebras because of multiplicities
appearing in irreducible representations decomposition. But as we will see
below, even in these cases some information about the full S-matrix can be
extracted from requirement of consistency of derived TBA equations. More
precisely, the requirement that the obtained TBA system will correspond to
the proper $Y$-system, fixes the crossing-unitarising prefactor of
corresponding S-matrices. In contrast, in the cases $A_{n}$ and $C_{n}$ the
irrep. decomposition of tensor product of any two fundamental
representations is known, and full S-matrix may be written explicitly.
Derivation of TBA $Y$-system is exact in the sense the assumption about this
prefactor can be checked exactly. In all the cases of the considered Lie
algebras $G$ we show that the derived TBA equations lead to the correct
ground state free energy reproducing in the UV limit the central charge of
the coset.

In the second section we show how results about transfer matrix
diagonalization may be used in the derivation of TBA equations in the
framework of Bethe Ansatz (BA) string hypothesis. We show the role of
magnonic degrees of freedom and their relation to the main massive
particles. In the third section we show that in the thermodynamic limit one
of the possible magnonic degrees of freedom is always ''frozen'' -- it
always has zero density of holes, and we perform the reduction of this
degree of freedom in the equations. The main fourth part is devoted to the
transformation of obtained system of equations to the form of $Y$-system. As
we show, this transformation is possible and requires a natural assumption
about the form of the full S-matrix. Using results of \cite{KN}, we show
that the obtained $Y$systems has correct high temperature behavior,
reproducing central charges of the coset for any $G=A_{n},B_{n},C_{n},D_{n}$%
. In the fifth section we show that the assumption made about the form of
the crossing unitarising scalar prefactor of the S-matrix, is really correct
in the $A_{n,}C_{n}$ cases. We also discuss particular cases of the general
S-matrix for some specific models with Yangian symmetry. We conclude the
paper by brief discussion. In Appendix we collect kernels and technical
details of the TBA derivation.

\section{Transfer matrix diagonalization and TBA equations}

In contrast to the case of relativistic two dimensional integrable quantum
field theories (IQFT) with elastic (diagonal) S-matrices, such as, e.g.
S-matrices of affine Toda field theories (see, e.g. \cite{BCDS}\cite{D}\cite
{DGZ}), the transfer matrix diagonalization for the IQFT with internal Lie
algebraic symmetry, is hard, and, in general, non solvable problem. More
progress was achieved in solution of this problem in lattice models, such as
spin chains or RSOS models invariant under some Lie algebraic symmetry,
rather in relativistic IQFT. Here we recall one simple method to attach the
lattice results, derived in RSOS like traditional Bethe ansatz methods, to
explicit derivation of TBA equations for S-matrices describing integrable
perturbations of $\frac{G_{l}\times G_{m}}{G_{l+m}}$ coset CFTs ($%
G=A_{n},B_{n},C_{n},D_{n}$) relevantly perturbed by operator $(1,1,adj)$.
The S-matrix for this IQFT was conjectured a long time ago \cite{ABL} and it
has the form
\begin{equation}
S_{ab}(\theta )=X_{ab}(\theta )S_{ab}^{(l)}(\theta )\otimes
S_{ab}^{(m)}(\theta ),  \label{sm}
\end{equation}
where $S_{ab}^{(k)}(\theta )$ is unitary, crossing symmetric ''minimal''
(without poles in the physical strip $0<Im\theta <\pi $) RSOS-like S-matrix
for scattering of two particles from two multiplets corresponding to
fundamental weights $a$ and $b$ of algebra $G$. $k$ is the restriction level
of RSOS model. $X_{ab}$ is a CDD factor which generates poles for the
S-matrix corresponding to each fundamental weight of $G$, and guarantee the
bootstrap closure. Recall that from the point of view of quantum groups, $%
S_{ab}^{(l)}(\theta )$ RSOS S-matrix has the $U_{q_{1}}(G^{(1)})$ symmetry
with $q$ a root of unity $q=-e^{-i\pi /(l+g)}$.

The procedure of TBA derivation is standard: we pull one particle $j$ from
the fundamental multiplet $a_{j}$ with rapidity $\theta _{j}$, through a gas
of other particles living on a circle of length $\mathcal{L}$. On the way it
scatters on each other particle $a_{i}$ with rapidity $\theta _{i}$ with the
S-matrix $S_{a_{j}a_{i}}(\theta _{j}-\theta _{i})$, giving rise to the
transfer matrix
\begin{equation}
T^{a_{j}}(\theta _{j}|\theta _{i_{1}},...,\theta _{i_{N}})=\prod_{i=1,\neq
j}^{\mathcal{N}}S_{a_{j}a_{i}}(\theta _{j}-\theta _{i}).
\end{equation}
The requirement of the wave function periodicity looks like
\begin{equation}
e^{im_{a_{j}}\mathcal{L}\sinh \theta _{j}}T^{a_{j}}(\theta _{j}|\theta
_{j+1},...,\theta _{\mathcal{N}},\theta _{1},...,\theta _{j-1})=1.
\label{tmeq1}
\end{equation}
Non diagonality of the scattering leads to the change of states of the
particle inside its multiplet. In terms of Bethe ansatz, this change is
taken into account by means of magnonic excitations described by Bethe
ansatz equation (BAE). They are responsible for the non diagonal part of the
S-matrix, defined by spectral decomposition, whereas the diagonal part is
defined by the prefactors before this spectral decomposition -- $X_{ab}$,
and crossing-unitarising prefactors $\sigma _{ab}^{(l,m)}$ of the minimal
S-matrices $S_{ab}^{(l,m)}$. Explicit and full form of the non diagonal part
of the transfer matrix in terms of magnonic degrees of freedom is
complicated. But it was proven by Kirillov and Reshetikhin for simply laced
algebras \cite{BR}, and conjectured, and partly proved, for non simply laced
algebras \cite{KNS}, that in the thermodynamic limit, when the number of
particles $\mathcal{N}$ together with the length $\mathcal{L}$ are going to
infinity, there is dominating ''top'' term for transfer matrix. Leaving only
this top term, we have for the transfer matrix the following expression
\begin{equation}
T^{a_{j}}(\theta _{j}|\theta _{i_{1}},...,\theta _{i_{N}})=\prod_{i=1,\neq
j}^{\mathcal{N}}X_{a_{j}a_{i}}(\theta _{j}-\theta _{i})\sigma
_{a_{j}a_{i}}^{(l)}(\theta _{j}-\theta _{i})\sigma
_{a_{j}a_{i}}^{(m)}(\theta _{j}-\theta _{i})\times  \label{tm1}
\end{equation}
\[
\times \prod_{\alpha =1}^{M_{a_{j}}^{(l)}}\frac{\sinh \left( \frac{\pi }{%
2(l+g)}\left( \theta _{j}-u_{\alpha }^{a_{j}}+it_{a_{j}}^{-1}\right) \right)
}{\sinh \left( \frac{\pi }{2(l+g)}\left( \theta _{j}-u_{\alpha
}^{a_{j}}-it_{a_{j}}^{-1}\right) \right) }\prod_{\alpha =1}^{M_{a_{j}}^{(m)}}%
\frac{\sinh \left( \frac{\pi }{2(m+g)}\left( \theta _{j}-v_{\alpha
}^{a_{j}}+it_{a_{j}}^{-1}\right) \right) }{\sinh \left( \frac{\pi }{2(m+g)}%
\left( \theta _{j}-v_{\alpha }^{a_{j}}-it_{a_{j}}^{-1}\right)
\right) }.
\]
Here the last line describes
the ''top'' term contribution, according to \cite{KNS}, $g$ is a
dual Coxeter number of algebra $G$, $t_{a}$ - integer
number related to the node $a$ of Dynkin diagram of $G$. For algebras $%
G=A_{n},B_{n},C_{n},D_{n}$ considered in this paper, it is equal to 1 for
long root nodes, and 2 -- for short root nodes. Sets of numbers $u_{\alpha
}^{a_{j}},v_{\alpha }^{a_{j}}$ satisfy the BAE \cite{RW}

\begin{equation}
\prod_{j=1}^{\mathcal{N}}\frac{\sinh \left( \frac{\pi }{2(l+g)}\left(
u_{\alpha }^{b}-\theta _{j}+i\omega _{a_{j}}\cdot \alpha _{b}\right) \right)
}{\sinh \left( \frac{\pi }{2(l+g)}\left( u_{\alpha }^{b}-\theta _{j}-i\omega
_{a_{j}}\cdot \alpha _{b}\right) \right) }=\Omega _{\alpha
}^{b}\prod_{c=1}^{n}\prod_{\beta =1}^{M_{c}^{(l)}}\frac{\sinh \left( \frac{%
\pi }{2(l+g)}\left( u_{\alpha }^{b}-u_{\beta }^{c}+i\alpha _{b}\cdot \alpha
_{c}\right) \right) }{\sinh \left( \frac{\pi }{2(l+g)}\left( u_{\alpha
}^{b}-u_{\beta }^{c}-i\alpha _{b}\cdot \alpha _{c}\right) \right) },
\label{BAE}
\end{equation}
and the same for $v_{\alpha }^{a}$, with $l$ replaced by $m$. Here $\omega
_{a}$,$\alpha _{a}$ are fundamental weights and simple roots of $G$. $\Omega
_{\alpha }^{b}$ is a constant which is not important for us here.

It is worth to notice here that rapidities of the physical particles $\theta
_{j}$ appear as inhomogeneities in the l.h.s. of the BAE (\ref{BAE}) for
magnonic degrees of freedom. This procedure differes from light cone lattice 
regularization scheme for relativistic IQFT, when
one considers BAE like (\ref{BAE}) with light cone inhomogeneities $\Theta $
in the l.h.s., and mass scale is introduced in a special scaling limit: $%
\Theta \rightarrow \infty $, lattice step $a\rightarrow 0$.

It is important that according to the general conjecture \cite{BR}\cite{KNS}
about the top term (\ref{tm1}), this transfer matrix eigenvalue is
associated to a representation of $G_{n}\times G_{n}$ with highest weight
\begin{equation}
(\mu ^{(l)},\mu ^{(m)})=\left( \sum_{i=1}^{\mathcal{N}}\omega
_{a_{i}}-\sum_{a=1}^{n}M_{a}^{(l)}\alpha _{a},\sum_{i=1}^{\mathcal{N}}\omega
_{a_{i}}-\sum_{a=1}^{n}M_{a}^{(m)}\alpha _{a}\right),  \label{hw}
\end{equation}
and the RSOS restrictions $\mu ^{(l)}\theta \leq l,\mu ^{(m)}\theta \leq m$
impose important restrictions on the possible values of $M_{a}^{(l,m)}$,
where $\theta $ is the highest root of $G$.

Procedure of taking the thermodynamic limit $\mathcal{N}\rightarrow \infty ,%
\mathcal{L}\rightarrow \infty $ is standard: rapidities $\theta _{j}$ become
dense, and solutions of BAE form strings. The string hypothesis for any
algebra $G$ was formulated in \cite{K} and it looks as follows. In the
thermodynamic limit macroscopically large amount of solutions of BAE have
the form of color $a$, $k$-strings
\begin{equation}
u_{\alpha }^{a}=u_{a,k}+it_{a}^{-1}(k+1-2j),\ j=1,...,k,  \label{str}
\end{equation}
where $u_{a,k}$ is real and has some density (density of string), and
possible values of $k$ are $k=1,...,t_{a}l$.

Before we start the thermodynamic calculation let us fix some notations. We
use the following Cartan matrices for $G$: $C_{ab}=\frac{2\alpha _{a}\alpha
_{b}}{\alpha _{a}\alpha _{a}}$. As we already defined, $t_{a}=\frac{2}{%
\alpha _{a}\alpha _{a}}$. The fundamental weights satisfy $\alpha _{a}\omega
_{b}=\delta _{ab}t_{a}^{-1}$. We also define $t_{ab}=\max (t_{a},t_{b})$. We
use the following Fourier transform convention
\[
f(u)=\frac{1}{2\pi }\int_{-\infty }^{\infty }\widehat{f}(\omega
)e^{iu\omega }d\omega ,\ \widehat{f}(\omega )=\int_{-\infty
}^{\infty }f(u)e^{-iu\omega }du.
\]

Now the standard thermodynamic calculation goes as follows (see, for
example, \cite{BR}). Consider the logarithm of (\ref{BAE}). The possible
ambiguity $2\pi iq$ can have a holes in its continuous occupation if integer
numbers $q$. In the thermodynamic limit such holes form hole densities. We
sum up these equations over $u_{\alpha }^{a}$ belonging to a color $a$ $k$%
-string, introducing densities $\rho _{k}^{a}$ for real coordinates of
strings $u_{a,k}$, and hole densities $\widetilde{\rho }_{k}^{a}$ for holes
in $1/\mathcal{L}$ normalized $q$ distributions. The same procedure we
perform with the BAE for $v_{\alpha }^{a}$, introducing magnonic densities $%
\eta _{k}^{a}$ for real coordinates of strings $v_{a,k}$, and densities of
holes $\widetilde{\eta }_{k}^{a}$. The same procedure can be applied to the $%
\ln $ of \ the eq. (\ref{tmeq1}) with explicit form of the transfer matrix
given by (\ref{tm1}), with introduction of particle $a$ densities $\sigma
^{a}$, and hole densities for them $\widetilde{\sigma }^{a}$. The resulting
equations have the form \cite{K}, which is the simplest after the passing to
the Fourier transform. Here and below we will work mostly in the $\omega $
space, so we will omit the hat on all the variables depending on $\omega $,
and moreover, will omit their argument $\omega $, except for the cases when
it will be different from $\omega $.
\begin{eqnarray}
\widetilde{\sigma }^{a} &=&\frac{m_{a}}{2\pi }\widehat{\cosh \theta }%
-\sum_{b=1}^{n}Y_{ab}\sigma ^{b}-\sum_{j=1}^{t_{a}l}a_{j/t_{a}}^{(l+g)}\rho
_{j}^{a}-\sum_{j=1}^{t_{a}m}a_{j/t_{a}}^{(m+g)}\eta _{j}^{a}  \label{m} \\
a_{j/t_{a}}^{(l+g)}\sigma ^{a} &=&\widetilde{\rho }_{j}^{a}+\sum_{b=1}^{n}%
\sum_{k=1}^{t_{b}l}M_{ab}A_{ab}^{(l+g)jk}\rho _{k}^{b}  \label{nm1} \\
a_{j/t_{a}}^{(m+g)}\sigma ^{a} &=&\widetilde{\eta }_{j}^{a}+\sum_{b=1}^{n}%
\sum_{k=1}^{t_{b}l}M_{ab}A_{ab}^{(m+g)jk}\eta _{k}^{b}.  \label{nm2}
\end{eqnarray}
Here $a=1,...,n$, $j=1,...,t_{a}l$ in eq. (\ref{nm1}), $j=1,...,t_{a}m$ in
eq. (\ref{nm2}). Masses $m_{a}$ of multiplets for different algebras $G$
will be listed later. $\widehat{\cosh \theta }$ is Fourier transform of $%
\cosh \theta $ function, and the kernels in the $\omega $ space look like
(here and in what follows we use the short notations for $\sinh $ and $\cosh
$ functions $\sinh (\omega x)=[x],\cosh (\omega x)=(x)$)
\begin{eqnarray}
a_{k}^{(L)} &=&\frac{[L-k]}{[L]},  \label{ker} \\
A_{ab}^{(L)jk} &=&\frac{[\min (jt_{a}^{-1},kt_{b}^{-1})][L-\max
(jt_{a}^{-1},kt_{b}^{-1})]}{[t_{ab}^{-1}][L]},  \nonumber \\
M_{ab} &=&C_{ab}\frac{t_{b}}{t_{ab}}+2\delta _{ab}\left(
(t_{a}^{-1})-1\right),  \nonumber
\end{eqnarray}
and $Y_{ab}$ is Fourier transform of $Y_{ab}(\theta )$ which comes from the
S-matrix $S_{ab}(\theta )$ prefactors
\begin{equation}
Y_{ab}(\theta )=\frac{1}{2\pi i}\frac{d}{d\theta }\ln \left( X_{ab}(\theta
)\sigma _{ab}^{(l)}(\theta )\sigma _{ab}^{(m)}(\theta )\right).  \label{y}
\end{equation}
The equations (\ref{m})-(\ref{nm2}) are the basic equations for the TBA
derivation. Before we come to it, one important step is necessary. As it was
firstly noted by Bazhanov and Reshetikhin, one of the degrees of freedom in
these equations is frozen. As we will see in the next section, oppositely to
spin models, the RSOS restriction dictates $\widetilde{\rho }_{t_{a}l}^{a}=%
\widetilde{\eta }_{t_{a}m}^{a}=0$ in the thermodynamic limit, which can be
used for reduction of the highest $t_{a}l$ and $t_{a}m$ strings.

\section{Maximal string reduction}

Consider the zero mode of the $l$-system (\ref{nm1}) for $j=t_{a}l$. Using
explicit form of the kernels (\ref{ker}), one has
\[
\frac{g}{l+g}\sigma ^{a}(0)=\widetilde{\rho }_{t_{a}l}^{a}(0)+\frac{g}{l+g}%
\sum_{b=1}^{n}C_{ab}\sum_{k=1}^{t_{b}l}k\rho _{k}^{b}(0).
\]
In the thermodynamic limit the highest weight (\ref{hw}), which dominates in
the transfer matrix eigenvalue, becomes
\[
\mu ^{(l)}=\mathcal{L}\sum_{a=1}^{n}\left( \sigma ^{a}(0)\omega
_{a}-\sum_{k=1}^{t_{a}l}k\rho _{k}^{a}(0)\alpha _{a}\right),
\]
and due to the previous equation, using $\sum_{b=1}^{n}C_{ab}\omega
_{b}=\alpha _{a}$, it gives
\[
\mu ^{(l)}=\mathcal{L}\frac{l+g}{g}\sum_{a=1}^{n}\widetilde{\rho }%
_{t_{a}l}^{a}(0)\omega _{a}.
\]
RSOS restriction $\mu ^{(l)}\theta \leq l$ now gives $\mathcal{L}\frac{l+g}{g%
}\sum_{a=1}^{n}\widetilde{\rho }_{t_{a}l}^{a}(0)\omega _{a}\theta \leq l$.
The fact that $\widetilde{\rho }_{t_{a}l}^{a}(0)$ may be only non negative,
and that $\omega _{a}\theta $ is a positive number for any $a$ and any $G$,
necessarily requires that $\widetilde{\rho }_{t_{a}l}^{a}(0)=0$ for each $a$
in $\mathcal{L}\rightarrow \infty $ limit. From this immediately follows
that $\widetilde{\rho }_{t_{a}l}^{a}(\omega )=0$ for any $\omega $. The same
is valid for $\widetilde{\eta }_{t_{a}l}^{a}(\omega )=0$.

Using this fact, we express $\rho _{t_{a}l}^{a}$ through other variables.
Eq. (\ref{nm1}) gives for $j=t_{a}l$%
\[
a_{l}^{(l+g)}\sigma
^{a}=\sum_{b=1}^{n}\sum_{k=1}^{t_{b}l-1}M_{ab}A_{ab}^{(l+g)t_{a}l,k}\rho
_{k}^{b}+\frac{\left[ l\right] \left[ g\right] }{\left[ l+g\right] }%
\sum_{b=1}^{n}\frac{M_{ab}}{\left[ t_{ab}^{-1}\right] }\rho
_{t_{b}l}^{b}.
\]
The inverse of the matrix $\frac{M_{ab}}{\left[ t_{ab}^{-1}\right] }$ we
will denote $\widetilde{A}_{ab}^{G_{n}}$. We will use also another matrix $%
A_{ab}^{G_{n}}=\frac{2(1)}{[1]}\widetilde{A}_{ab}^{G_{n}}$. The inverse of
the matrices $A_{ab}^{G_{n}}$ will be called $K_{ab}^{G_{n}}=\left(
A^{G_{n}}\right) _{ab}^{-1}$. $\widetilde{A}^{G_{n}}$ can be calculated case
by case for each of the four algebras we consider here. The list of matrices
$\widetilde{A}_{ab}^{G_{n}}$ for each $G$ one can find in the Appendix. So
we have
\begin{equation}
\rho _{t_{a}l}^{a}=\frac{1}{\left[ l\right] }\sum_{b=1}^{n}\widetilde{A}%
_{ab}^{G_{n}}\sigma ^{b}-\sum_{k=1}^{t_{a}l-1}a_{k/t_{a}}^{(l)}\rho _{k}^{a}.
\label{romax}
\end{equation}
In the same way one gets
\begin{equation}
\eta _{t_{a}m}^{a}=\frac{1}{\left[ m\right] }\sum_{b=1}^{n}\widetilde{A}%
_{ab}^{G_{n}}\sigma ^{b}-\sum_{k=1}^{t_{a}m-1}a_{k/t_{a}}^{(l)}\eta _{k}^{a}.
\label{etamax}
\end{equation}
Substitution of (\ref{romax}) and (\ref{etamax}) into (\ref{nm1}) and (\ref
{nm2}) gives, after some simple algebra, reduced magnonic BAE:
\begin{eqnarray}
a_{j/t_{a}}^{(l)}\sigma ^{a} &=&\widetilde{\rho }_{j}^{a}+\sum_{b=1}^{n}%
\sum_{k=1}^{t_{b}l-1}M_{ab}A_{ab}^{(l)jk}\rho _{k}^{b}  \label{nmr1} \\
a_{j/t_{a}}^{(m)}\sigma ^{a} &=&\widetilde{\eta }_{j}^{a}+\sum_{b=1}^{n}%
\sum_{k=1}^{t_{b}l-1}M_{ab}A_{ab}^{(m)jk}\eta _{k}^{b}.  \label{nmr2}
\end{eqnarray}

Before substitution of (\ref{romax}) and (\ref{etamax}) into the massive
equation (\ref{m}) we will make an important \textbf{assumption}: for all
the algebras $G$ the kernel $Y_{ab}$ has the form
\begin{equation}
Y_{ab}=\varphi A_{ab}^{G_{n}},  \label{asum}
\end{equation}
with some scalar function $\varphi (\omega )$. As we will see later, this
assumption is correct in both $G=A_{n}$ and $G=C_{n}$ cases, when the full
S-matrix, including its spectral decomposition, is known. With this
assumption substitution gives
\begin{equation}
\widetilde{\sigma }^{a}=\frac{m_{a}}{2\pi }\widehat{\cosh \theta }-%
\widetilde{\varphi }\sum_{b=1}^{n}A_{ab}^{G_{n}}\sigma
^{b}-\sum_{j=1}^{t_{a}l-1}a_{j/t_{a}}^{(l)}\rho
_{j}^{a}-\sum_{j=1}^{t_{a}m-1}a_{j/t_{a}}^{(m)}\eta _{j}^{a},  \label{mr}
\end{equation}
where
\begin{equation}
\widetilde{\varphi }=\varphi +a_{l}^{(l+g)}\frac{\left[ 1\right] }{\left[ l%
\right] 2(1)}+a_{m}^{(m+g)}\frac{\left[ 1\right] }{\left[ m\right] 2(1)}.
\label{phi1}
\end{equation}
So, effective reduction of the thermodynamically ''frozen'' degrees of
freedom leads to the system of equations (\ref{mr}),(\ref{nmr1}) and (\ref
{nmr2}), very similar to the original ones. The effect of reduction is the
change of parameters $l+g\rightarrow l,\left( m+g\rightarrow m\right) $ and
possible string length is now not grater then $t_{a}l-1,\left(
t_{a}m-1\right) $.

\section{Thermodynamic and transformation to $Y$-system}

In the massive BA equation (\ref{mr}) we have magnonic degrees of freedom --
magnonic densities, which represent not excitations but rather Dirac vacuum
for magnonic degrees of freedom. If we want the massive equation to be
explicitly dependent on magnonic excitations, represented by hole densities $%
\widetilde{\rho }_{j}^{a}$,$\widetilde{\eta }_{j}^{a}$, we should solve eqs.
(\ref{nmr1}) and (\ref{nmr2}) with respect to $\rho _{j}^{a}$,$\eta _{j}^{a}$
as functions of $\widetilde{\rho }_{j}^{a}$,$\widetilde{\eta }_{j}^{a}$ and $%
\sigma ^{a}$, and substitute these solutions into the eq.(\ref{mr}). As we
will see on the stage of doing thermodynamics, this will give us a
possibility to convert the system (\ref{mr}),(\ref{nmr1}),(\ref{nmr2}) more
easily to the form close to a so called $Y$-system. We have it as a goal,
since such kind of systems were classified \cite{KN} according to affine
symmetry standing behind the system, and this classification adjust to each
such system its UV limit (in thermodynamic terms, the limit $T\rightarrow
\infty $ ). This limit contains information about the central charge of the
unperturbed CFT -- the coset $\frac{G_{l}\times G_{m}}{G_{l+m}}$. Actually,
we don't need the expressions for $\rho _{j}^{a}$,$\eta _{j}^{a}$
themselves, but only the specific combinations entering the massive
equation, like $\sum_{j=1}^{t_{a}l-1}a_{j/t_{a}}^{(l)}\rho _{j}^{a}$.

\subsection{Transformation of magnonic degrees of freedom and S-matrix
prefactor}

General description of the procedure is possible, but we will do calculation
separately for both of non simply laced algebras, since this will make the
formulas more transparent. We multiply eq. (\ref{nmr1}) by the matrix
\begin{equation}
K_{a}^{kj}=\delta _{kj}+\frac{1}{2(t_{a}^{-1})}\left(
C_{kj}^{A_{t_{a}l-1}}-2\delta _{kj}\right) =K^{kj}(t_{a}^{-1}),  \label{kdef}
\end{equation}
and sum over $j$ from 1 to $t_{a}l-1$. Using the identity
\begin{equation}
\sum_{j=1}^{t_{a}l-1}K_{a}^{kj}a_{j/t_{a}}^{(l)}=\frac{\delta _{k1}}{%
2(t_{a}^{-1})},  \label{id1}
\end{equation}
one gets the equation
\begin{equation}
\frac{\delta _{k1}}{2(t_{a}^{-1})}\sigma ^{a}=\sum_{j=1}^{t_{a}l-1}K_{a}^{kj}%
\widetilde{\rho }_{j}^{a}+\sum_{b=1}^{n}\sum_{j=1}^{t_{b}l-1}J_{ab}^{kj}\rho
_{j}^{b}.  \label{nmrj}
\end{equation}
After some algebra the kernel $J$ may be written in the universal form:
\begin{eqnarray}
J_{ab}^{kj} &=&\frac{M_{ab}}{2(t_{a}^{-1})\left[ t_{ab}^{-1}\right] }\left( %
\left[ t_{a}^{-1}\right] \delta _{t_{b}k,t_{a}j}+\right.  \label{j} \\
&&\left. +\sum_{q=1}^{t_{b}/t_{a}-1}\left[ qt_{b}^{-1}\right]
\left( \delta _{t_{b}(k+1)-t_{a}q,t_{a}j}+\delta
_{t_{b}(k-1)+t_{a}q,t_{a}j}\right) \right).  \nonumber
\end{eqnarray}

\textbf{Simply laced cases (}$\mathbf{A}_{n}\mathbf{,D}_{n}$\textbf{)}.

In these cases all $t_{a}=1$, and $J$ has a simple form
\[
J_{ab}^{kj}=\frac{M_{ab}}{2(1)}\delta _{kj},
\]
and eq. (\ref{nmrj}) has the form
\begin{equation}
\frac{\delta _{k1}}{2(1)}\sigma ^{a}=\sum_{j=1}^{l-1}K^{kj}(\omega )%
\widetilde{\rho }_{j}^{a}+\sum_{b=1}^{n}K_{ab}^{G_{n}}\rho _{j}^{b}.
\label{nmsl}
\end{equation}
Multiplying it by $a_{k}^{(l)}$ with summation $\sum_{k=1}^{l-1}$, and using
the identity (\ref{id1}), one gets the equation for variable $%
x^{a}=\sum_{k=1}^{l-1}a_{k}^{(l)}\rho _{k}^{a},a=1,...,n$%
\[
\sum_{b=1}^{n}M_{ab}x^{b}=a_{1}^{(l)}\sigma ^{a}-\widetilde{\rho
}_{1}^{a},
\]
which one can easily solve using the inverse kernels $A^{G_{n}}$ (\ref{a})
or (\ref{d}):
\begin{equation}
x^{a}=\frac{1}{2(1)}\sum_{b=1}^{n}A_{ab}^{G_{n}}\left( a_{1}^{(l)}\sigma
^{a}-\widetilde{\rho }_{1}^{a}\right).  \label{xa}
\end{equation}

\textbf{Non simply laced cases.}

Unfortunately, the calculations in non simply laced cases are technically
more involved, although straigtforward. We decided to present them in
details. When some of $t_{a}$ are equal to 2, one has from (\ref{j}) the
following form of $J$:

\begin{itemize}
\item  when $t_{a}=t_{b}=1$,
\[
J_{ab}^{kj}=\frac{M_{ab}}{2(1)}\delta _{kj},
\]

\item  when $t_{a}=1,t_{b}=2$,
\[
J_{ab}^{kj}=\frac{\left( \frac{1}{2}\right) }{\left( 1\right)
}M_{ab}\left( \delta _{k,\frac{j}{2}}+\frac{1}{2\left(
\frac{1}{2}\right) }\left( \delta _{k,\frac{j-1}{2}}+\delta
_{k,\frac{j+1}{2}}\right) \right),
\]

\item  when $t_{a}=2,t_{b}=1$,
\[
J_{ab}^{kj}=\frac{M_{ab}}{2\left( \frac{1}{2}\right) }\delta _{\frac{k}{2}%
,j},
\]

\item  when $t_{a}=2,t_{b}=2$,
\[
J_{ab}^{kj}=\frac{M_{ab}}{2\left( \frac{1}{2}\right) }\delta
_{kj}.
\]
\end{itemize}

\textbf{Case }$\mathbf{C}_{n}$\textbf{. }

Equations (\ref{nmrj}) can be written as
\begin{eqnarray}
\frac{\delta _{j1}}{2(1/2)}\sigma ^{a} &=&\sum_{k=1}^{2l-1}K^{jk}\left(
\frac{\omega }{2}\right) \widetilde{\rho }_{k}^{a}+\frac{1}{2(1/2)}%
\sum_{b=1}^{n-1}M_{ab}\rho _{j}^{b}+\frac{1}{2(1/2)}M_{an}\rho _{j/2}^{n} \nonumber\\
\frac{\delta _{j1}}{2(1)}\sigma ^{n} &=&\sum_{k=1}^{l-1}K^{jk}\left( \omega
\right) \widetilde{\rho }_{k}^{n}+\frac{(1/2)}{(1)}M_{n,n-1}%
\sum_{k=1}^{2l-1}G^{jk}\rho _{k}^{n-1}+\frac{1}{2(1)}M_{nn}\rho
_{j}^{n}. \nonumber
\end{eqnarray}
Here in the first equation $a=1,...,n-1,j=1,...,2l-1$, and in the second -- $%
j=1,...,l-1$, and the kernel
\begin{equation}
G^{jk}=\delta _{j,k/2}+\frac{1}{2(1/2)}\left( \delta _{j,\frac{k+1}{2}%
}+\delta _{j,\frac{k-1}{2}}\right).  \label{g}
\end{equation}
In these equations and below a fractional index at any variable $\rho ,%
\widetilde{\rho },\eta ,\widetilde{\eta }$ means that it is equal to zero.
We would like to rewrite these equations using $K$ kernels instead of $M$
(see Appendix):
\begin{eqnarray}
\frac{\delta _{j1}}{2(1/2)}\sigma ^{a} &=&\sum_{k=1}^{2l-1}K^{jk}\left(
\frac{\omega }{2}\right) \widetilde{\rho }_{k}^{a}+\frac{1}{f}%
\sum_{b=1}^{n-1}K_{ab}^{C_{n}}\rho _{j}^{b}+\frac{1}{f}K_{an}^{C_{n}}\rho
_{j/2}^{n}  \label{c1} \\
\frac{\delta _{j1}}{2(1)}\sigma ^{n} &=&\sum_{k=1}^{l-1}K^{jk}\left( \omega
\right) \widetilde{\rho }_{k}^{n}+\sum_{b=1}^{n-1}%
\sum_{k=1}^{2l-1}G^{jk}K_{nb}^{C_{n}}\rho _{k}^{b}+\rho _{j}^{n}.  \label{c2}
\end{eqnarray}

Multiplying the first equation by $a_{j/2}^{(l)}$ and taking the sum $%
\sum_{j=1}^{2l-1}$, and the second -- by $a_{j}^{(l)}$ and taking the sum $%
\sum_{j=1}^{l-1}$, we use the identity
\[
\sum_{j=1}^{l-1}a_{j}^{(l)}G^{jk}=a_{k/2}^{(l)}-\frac{\delta
_{k1}}{2(1/2)}
\]
and explicit form of $M$. We obtain the following system of equations
\begin{eqnarray}
\frac{f}{2(1/2)}a_{1/2}^{(l)}\sigma ^{a} &=&\frac{f}{2(1/2)}\widetilde{\rho }%
_{1}^{a}+\sum_{b=1}^{n-1}K_{ab}^{C_{n}}x^{b}+K_{an}^{C_{n}}x^{n}
\label{tmc1} \\
\frac{1}{2(1)}a_{1}^{(l)}\sigma ^{n} &=&\frac{1}{2(1)}\widetilde{\rho }%
_{1}^{n}-\frac{(1/2)}{(1)}x^{n-1}+x^{n}+\frac{1}{2(1)}\rho _{1}^{n-1}
\label{tmc2}
\end{eqnarray}
for the variables $x^{a}=\sum_{j=1}^{2l-1}a_{j/2}^{(l)}\rho
_{j}^{a},a=1,...n-1$, and $x^{n}=\sum_{j=1}^{l-1}a_{j}^{(l)}\rho _{j}^{a}$.
Their solution (for details see Appendix) has the form
\begin{eqnarray}
x^{n} &=&\frac{1}{2(1)}\left[ A_{nn}^{C_{n}}\left( a_{1}^{(l)}\sigma ^{n}-%
\widetilde{\rho }_{1}^{n}\right) +\sum_{b=1}^{n-1}\left( 2\left( 1/2\right)
A_{nb}^{C_{n}}\left( a_{1/2}^{(l)}\sigma ^{b}-\widetilde{\rho }%
_{1}^{b}\right) -\right. \right.  \nonumber \\
&&\left. \left. -A_{nb}^{C_{n}}\left( \sigma
^{b}-2(1/2)\sum_{k=1}^{2l-1}K_{1k}(\omega /2)\widetilde{\rho }%
_{k}^{b}\right) \right) \right],  \label{xnc}
\end{eqnarray}
\begin{eqnarray}
x^{a} &=&\frac{(1/2)}{(1)}\left( \frac{A_{na}^{C_{n}}}{2(1/2)}\left(
a_{1}^{(l)}\sigma ^{n}-\widetilde{\rho }_{1}^{n}\right)
+\sum_{b=1}^{n-1}\left( a_{1/2}^{(l)}A_{ab}^{C_{n}}\sigma ^{b}-A_{ab}^{C_{n}}%
\widetilde{\rho }_{1}^{b}\right) \right) -  \nonumber \\
&&-A_{an}^{C_{n}}\left( A_{nn}^{C_{n}}\right) ^{-1}A_{nb}^{C_{n}}\left(
\frac{1}{2(1)}\sigma ^{b}-\frac{(1/2)}{(1)}\sum_{k=1}^{2l-1}K_{1k}(\omega /2)%
\widetilde{\rho }_{k}^{b}\right).  \label{xac}
\end{eqnarray}

\textbf{Case }$\mathbf{B}_{n}$\textbf{.}

In a similar way one can deal with the $B_{n}$ case. Eqs. (\ref{nmrj}) in
this case take the form
\begin{eqnarray}
\frac{\delta _{j1}}{2(1)}\sigma ^{a} &=&\sum_{k=1}^{l-1}K^{jk}\left( \omega
\right) \widetilde{\rho }_{k}^{a}+\frac{1}{2(1)}\sum_{b=1}^{n}M_{ab}\rho
_{j}^{b},\;a=1,...n-2 \nonumber\\
\frac{\delta _{j1}}{2(1)}\sigma ^{n-1} &=&\sum_{k=1}^{l-1}K^{jk}\left(
\omega \right) \widetilde{\rho }_{k}^{n-1}+\frac{1}{2(1)}\left(
M_{n-1,n-2}\rho _{j}^{n-2}+M_{n-1,n-1}\rho _{j}^{n-1}\right) + \nonumber\\
&&+\frac{(1/2)}{(1)}M_{n-1,n}\sum_{k=1}^{2l-1}G^{jk}\rho _{j}^{n} \nonumber\\
\frac{\delta _{j1}}{2(1/2)}\sigma ^{n}
&=&\sum_{k=1}^{2l-1}K^{jk}\left( \omega /2\right) \widetilde{\rho
}_{k}^{n}+\frac{1}{2(1/2)}M_{n,n-1}\rho
_{j/2}^{n-1}+\frac{1}{2(1/2)}M_{nn}\rho _{j}^{n}, \nonumber
\end{eqnarray}
where $j=1,...l-1$ in the first two equations, and $j=1,...2l-1$ - in the
last one. $G$ is the same as in eq. (\ref{g}). In terms of kernels $%
K^{B_{n}} $ (see Appendix), these equations look like

\begin{eqnarray}
\frac{\delta _{k1}}{2(1)}\sigma ^{a} &=&\sum_{j=1}^{l-1}K^{kj}\left( \omega
\right) \widetilde{\rho }_{j}^{a}+\sum_{b=1}^{n}K_{ab}^{B_{n}}\rho
_{k}^{b},\;a=1,...n-2  \label{b1} \\
\frac{\delta _{k1}}{2(1)}\sigma ^{n-1} &=&\sum_{k=1}^{l-1}K^{kj}\left(
\omega \right) \widetilde{\rho }_{j}^{n-1}+\sum_{b=1}^{n-1}K_{n-1b}^{B_{n}}%
\rho _{k}^{b}+K_{n-1n}^{B_{n}}\sum_{k=1}^{2l-1}G^{kj}\rho _{j}^{n}
\label{b2} \\
\frac{\delta _{k1}}{2(1/2)}\sigma ^{n} &=&\sum_{k=1}^{2l-1}K^{kj}\left(
\omega /2\right) \widetilde{\rho }_{j}^{n}+\frac{1}{f}K_{n,n-1}^{B_{n}}\rho
_{k/2}^{n-1}+\rho _{k}^{n}.  \label{b3}
\end{eqnarray}
Multiplying (\ref{b1}), (\ref{b2}) by $a_{k}^{(l)}$ with summation $%
\sum_{k=1}^{l-1}$, and (\ref{b3}) -- by $a_{k/2}^{(l)}$ with summation $%
\sum_{k=1}^{2l-1}$, as in the $C_{n}$ case, we get the following system of
linear equations
\begin{eqnarray}
\frac{a_{1}^{(l)}}{2(1)}\sigma ^{a} &=&\frac{1}{2(1)}\widetilde{\rho }%
_{1}^{a}+\sum_{b=1}^{n}K_{ab}^{B_{n}}x^{b},\;a=1,...n-2  \label{tmb1} \\
\frac{a_{1}^{(l)}}{2(1)}\sigma ^{n-1} &=&\frac{1}{2(1)}\widetilde{\rho }%
_{1}^{n-1}+\sum_{b=1}^{n-1}K_{n-1b}^{B_{n}}x^{b}+K_{n-1n}^{B_{n}}x^{n}-\frac{%
1}{2(1/2)}K_{n-1n}^{B_{n}}\rho _{1}^{n}  \label{tmb2} \\
\frac{a_{1/2}^{(l)}}{2(1/2)}\sigma ^{n} &=&\frac{1}{2(1/2)}\widetilde{\rho }%
_{1}^{n}+\frac{1}{f}K_{nn-1}^{B_{n}}x^{n-1}+x^{n}  \label{tmb3}
\end{eqnarray}
with respect to the indeterminates $x^{a}=\sum_{j=1}^{l-1}a_{j}^{(l)}\rho
_{j}^{a},a=1,...n-1$, and $x^{n}=\sum_{j=1}^{2l-1}a_{j/2}^{(l)}\rho _{j}^{a}$%
. Their solution (see Appendix) is
\begin{eqnarray}
x^{a} &=&\sum_{b=1}^{n-1}\frac{A_{ab}^{B_{n}}}{2(1)}\left( a_{1}^{(l)}\sigma
^{b}-\widetilde{\rho }_{1}^{b}\right) +\frac{(1/2)}{(1)}A_{an}^{B_{n}}\left(
a_{1/2}^{(l)}\sigma ^{n}-\widetilde{\rho }_{1}^{n}\right) -\frac{(1/2)}{(1)}%
A_{an}^{B_{n}}\rho _{1}^{n},  \label{xab} \\
x^{n} &=&\sum_{b=1}^{n-1}\frac{A_{nb}^{B_{n}}}{2(1)}\left( a_{1}^{(l)}\sigma
^{b}-\widetilde{\rho }_{1}^{b}\right) +\frac{(1/2)}{(1)}A_{nn}^{B_{n}}\left(
a_{1/2}^{(l)}\sigma ^{n}-\widetilde{\rho }_{1}^{n}\right) -\frac{%
A_{nn-1}^{B_{n}}}{2(1)}\rho _{1}^{n}.  \label{xnb}
\end{eqnarray}

The same solutions $y^{a},y^{n}$ one obtains for the other ($\eta $)
magnonic contributions to the massive equation (\ref{mr}). They have the
same form with change $\rho \rightarrow \eta ,\widetilde{\rho }\rightarrow
\widetilde{\eta },l\rightarrow m$

\bigskip

Now we substitute the obtained expressions for magnonic transfer matrix
contributions (\ref{xa}),(\ref{xac}),(\ref{xnc}),(\ref{xab}),(\ref{xnb})
into the massive equations (\ref{mr}). The non simply laced cases will again
be considered separately, but we start from the simply laced cases.

\textbf{Simply laced cases (}$\mathbf{A}_{n}\mathbf{,D}_{n}$\textbf{)}.

Substitution of (\ref{xa}), and the similar expression for the $\eta $
dependent part, into (\ref{mr}) gives
\[
\widetilde{\sigma }^{a}=\frac{m_{a}}{2\pi }\widehat{\cosh \theta }-\left(
\widetilde{\varphi }+\frac{a_{1}^{(l)}+a_{1}^{(m)}}{2(1)}\right)
\sum_{b=1}^{n}A_{ab}^{G_{n}}\sigma ^{b}+\frac{1}{2(1)}%
\sum_{b=1}^{n}A_{ab}^{G_{n}}\left( \widetilde{\rho
}_{1}^{b}+\widetilde{\eta }_{1}^{b}\right).
\]
Multiplication of this equation by the matrix $K^{G_{n}}$, the inverse one
to $A_{ab}^{G_{n}}$, gives the equation
\begin{equation}
\sum_{b=1}^{n}K_{ab}^{G_{n}}\widetilde{\sigma }^{b}=-\left\{ \left(
\widetilde{\varphi }+\frac{a_{1}^{(l)}+a_{1}^{(m)}}{2(1)}\right) \sigma ^{a}-%
\frac{\widetilde{\rho }_{1}^{a}+\widetilde{\eta }_{1}^{a}}{2(1)}\right\}.
\label{msl}
\end{equation}

The main feature of the mass spectrums $m_{a}$ for all the algebras, listed
in the Appendix, is that the vector $\frac{m_{a}}{2\pi }\widehat{\cosh
\theta }$ is an eigenvector of $K_{ab}^{G_{n}}$ with zero eigenvalue, which
leads to disappearing of this term in the last equation. It makes possible
to transform the system to so called universal form, when it does not
contain any other external functions or parameters, but only the variables
themselves (see below).

Together with (\ref{phi1}) the coefficient before $\sigma ^{a}$ in the last
equation may be written as
\[
\varphi +\frac{1}{2(1)}\left( \frac{\left[ 1\right] }{\left[ l\right] }%
a_{l}^{(l+g)}+a_{1}^{(l)}+\frac{\left[ 1\right] }{\left[ m\right] }%
a_{m}^{(m+g)}+a_{1}^{(m)}\right),
\]
which after some algebra can be written as
\[
\varphi +\frac{a_{1}^{(l+g)}+a_{1}^{(m+g)}}{2(1)}.
\]
The main conjecture is that
\begin{equation}
\varphi =\left( A_{g+l,g+l}^{A_{2g+l+m-1}}\right) ^{-1}=\left(
A_{g+m,g+m}^{A_{2g+l+m-1}}\right) ^{-1}.  \label{phi}
\end{equation}
As we will see in the next section, this is exactly what one has from the
S-matrix in the $A_{n}$ case, and we suppose the same expression is correct
for $D_{n}$ case too. This assumption will be shown correct for $%
S_{11}^{D_{n}}$. With this expression for $\varphi $, after some algebra,
one gets the coefficient before $\sigma ^{a}$ equal to 1.

\textbf{Non simply laced cases.}

As we will see now, this property of corresponding coefficient will remain
valid in non simply laced cases too, and will define the form of S-matrix
prefactor $\varphi $.

$\mathbf{C}_{n}$\textbf{\ case}.

Equations (\ref{mr}) written separately for $a\leq n-1$ and $a=n$, look like
\begin{eqnarray}
\widetilde{\sigma }^{a} &=&\frac{m_{a}}{2\pi }\widehat{\cosh \theta }-%
\widetilde{\varphi }\sum_{b=1}^{n}A_{ab}^{C_{n}}\sigma
^{b}-x^{a}-y^{a},\;a\leq n-1, \nonumber\\
\widetilde{\sigma }^{n} &=&\frac{m_{n}}{2\pi }\widehat{\cosh \theta }-%
\widetilde{\varphi }\sum_{b=1}^{n}A_{nb}^{C_{n}}\sigma
^{b}-x^{n}-y^{n}. \nonumber
\end{eqnarray}
Substitution of \ (\ref{xnc}) and (\ref{xac}) into them, after some algebra,
using $\sum_{b=1}^{n}K_{ab}^{C_{n}}A_{bc}^{C_{n}}=\delta _{ac}$, gives the
following equations
\begin{eqnarray}
\sum_{b=1}^{n}K_{ab}^{C_{n}}\widetilde{\sigma }^{b} &=&-\widetilde{\varphi }%
\sigma ^{a}-\frac{(1/2)}{(1)}\left( a_{1/2}^{(l)}+a_{1/2}^{(m)}\right)
\sigma ^{a}+\frac{(1/2)}{(1)}\left( \widetilde{\rho }_{1}^{a}+\widetilde{%
\eta }_{1}^{a}\right) \nonumber\\
\sum_{b=1}^{n}K_{nb}^{C_{n}}\widetilde{\sigma }^{b} &=&-\widetilde{\varphi }%
\sigma ^{n}-\frac{a_{1}^{(l)}+a_{1}^{(m)}}{2(1)}\sigma ^{n}+\frac{\widetilde{%
\rho }_{1}^{n}+\widetilde{\eta }_{1}^{n}}{2(1)}+ \nonumber\\
&&+\sum_{b=1}^{n-1}\frac{A_{nb}^{C_{n}}}{A_{nn}^{C_{n}}}\left( \frac{1}{2(1)}%
\sigma ^{b}-\frac{(1/2)}{(1)}\sum_{k=1}^{2l-1}K_{1k}(\omega /2)\widetilde{%
\rho }_{k}^{b}\right) \nonumber\\
+\{\widetilde{\rho } &\rightarrow &\widetilde{\eta },l\rightarrow
m\}. \nonumber
\end{eqnarray}

We recall that massive terms disappear since they are eigenvectors of $%
K^{C_{n}}$ with zero eigenvalue. Using the eq.(\ref{c1}) at $j=1$ and
explicit form of the kernels, one can see that the last sum in the last
equation is nothing but $\frac{1}{2(1)}\rho _{1}^{n-1}$, which gives the
system
\begin{eqnarray}
\sum_{b=1}^{n}K_{ab}^{C_{n}}\widetilde{\sigma }^{b} &=&-\left( \widetilde{%
\varphi }+\frac{(1/2)}{(1)}\left( a_{1/2}^{(l)}+a_{1/2}^{(m)}\right) \right)
\sigma ^{a}+\frac{(1/2)}{(1)}\left( \widetilde{\rho }_{1}^{a}+\widetilde{%
\eta }_{1}^{a}\right)  \label{kca} \\
\sum_{b=1}^{n}K_{nb}^{C_{n}}\widetilde{\sigma }^{b} &=&-\left( \widetilde{%
\varphi }+\frac{a_{1}^{(l)}+a_{1}^{(m)}}{2(1)}\right) \sigma ^{n}+\frac{%
\widetilde{\rho }_{1}^{n}+\widetilde{\eta }_{1}^{n}}{2(1)}+\frac{\rho
_{1}^{n-1}+\eta _{1}^{n-1}}{2(1)}  \label{kcn}
\end{eqnarray}
We make here the same conjecture as in the simply laced cases, which will be
checked by explicit calculation from the S-matrix in the next section
\[
\varphi =\left( A_{g+l,g+l}^{A_{2g+l+m-1}}\right) ^{-1}.
\]
(Recall that for $C_{n}$ $g=n+1$). As we saw above, this form of $\varphi $
leads to the fact that the coefficient before $\sigma ^{n}$ in eq. (\ref{kcn}%
) is 1. Using this fact, one can easily see that the coefficient before $%
\sigma ^{a}$ in eq. (\ref{kca}) is equal to $f=\frac{2(1/2)^{2}}{(1)}$. So
the final form of equations in the $C_{n}$ case is
\begin{eqnarray}
\frac{1}{f}\sum_{b=1}^{n}K_{ab}^{C_{n}}\widetilde{\sigma }^{b} &=&-\sigma
^{a}+\frac{\widetilde{\rho }_{1}^{a}+\widetilde{\eta }_{1}^{a}}{2(1/2)}
\label{kca1} \\
\sum_{b=1}^{n}K_{nb}^{C_{n}}\widetilde{\sigma }^{b} &=&-\sigma ^{n}+\frac{%
\widetilde{\rho }_{1}^{n}+\widetilde{\eta }_{1}^{n}}{2(1)}+\frac{\rho
_{1}^{n-1}+\eta _{1}^{n-1}}{2(1)}.  \label{kcn1}
\end{eqnarray}

$\mathbf{B}_{n}$ \textbf{case.}

Doing the same calculation as in the $C_{n}$ case, using (\ref{xab}) (\ref
{xnb}) and the identity
\[
1+\frac{(1/2)}{(1)}A_{n,n-1}^{B_{n}}=fA_{nn}^{B_{n}},
\]
one gets
\begin{eqnarray}
\sum_{b=1}^{n}K_{ab}^{B_{n}}\widetilde{\sigma }^{b} &=&-\left( \widetilde{%
\varphi }+\frac{a_{1}^{(l)}+a_{1}^{(m)}}{2(1)}\right) \sigma ^{a}+\frac{%
\widetilde{\rho }_{1}^{a}+\widetilde{\eta }_{1}^{a}}{2(1)}+\delta _{a,n-1}%
\frac{\rho _{1}^{a}+\eta _{1}^{a}}{2(1)}  \label{kba} \\
\sum_{b=1}^{n}K_{nb}^{B_{n}}\widetilde{\sigma }^{b} &=&-\left( \widetilde{%
\varphi }+\frac{(1/2)}{(1)}\left( a_{1/2}^{(l)}+a_{1/2}^{(m)}\right) \right)
\sigma ^{n}+\frac{(1/2)}{(1)}\left( \widetilde{\rho }_{1}^{a}+\widetilde{%
\eta }_{1}^{a}\right),  \label{kbn}
\end{eqnarray}
where in the first equation $a=1,...,n-1$.

The conjecture about the form of $\varphi $ (\ref{phi}) remains unchanged,
which gives, as we saw, coefficient 1 before $\sigma ^{a}$ in (\ref{kba}),
and coefficient $f$ before $\sigma ^{n}$ in (\ref{kbn}). It gives the
following final form of equations in the $B_{n}$ case
\begin{eqnarray}
\sum_{b=1}^{n}K_{ab}^{B_{n}}\widetilde{\sigma }^{b} &=&-\sigma ^{a}+\frac{%
\widetilde{\rho }_{1}^{a}+\widetilde{\eta }_{1}^{a}}{2(1)}+\delta _{a,n-1}%
\frac{\rho _{1}^{a}+\eta _{1}^{a}}{2(1)}  \label{kba1} \\
\frac{1}{f}\sum_{b=1}^{n}K_{nb}^{B_{n}}\widetilde{\sigma }^{b} &=&-\sigma
^{n}+\frac{\widetilde{\rho }_{1}^{a}+\widetilde{\eta }_{1}^{a}}{2(1/2)}.
\label{kbn1}
\end{eqnarray}

We see that the equations we got has a compact form, and are similar in
simply laced and non simply laced cases. Their universality is in particular
expressed by the fact that they don't contain mass terms.

Before we will do thermodynamic of the system we prefer to rewrite magnonic
and massive equations as one equation. It can be done if we introduce the
following new notations
\begin{eqnarray}
s_{j}^{a} &=&\widetilde{\rho }_{j}^{a},\;\widetilde{s}_{j}^{a}=\rho
_{j}^{a},\;j=1,...,t_{a}l-1, \nonumber\\
s_{j}^{a} &=&\widetilde{\eta }_{j}^{a},\;\widetilde{s}_{j}^{a}=\eta
_{j}^{a},\;j=-1,...,-t_{a}l+1, \nonumber\\
s_{0}^{a} &=&\sigma ^{a},\;\widetilde{s}_{0}^{a}=\widetilde{\sigma
}^{a}. \nonumber
\end{eqnarray}
As one can see, in terms of variables $s,\widetilde{s}$, the role of holes
and pseudoparticles flipped for magnonic degrees of freedom and remained the
same for massive ones. One can see that in terms of $s,\widetilde{s}$,
\textbf{simply laced} equations (\ref{msl}) and (\ref{nmsl}) together with
the copy of (\ref{nmsl}) for $\eta ,\widetilde{\eta }$, can be written as
one equation
\begin{equation}
\sum_{b=1}^{n}K_{ab}^{G_{n}}\widetilde{s}_{j}^{b}=-%
\sum_{j=-m+1}^{l-1}K^{kj}(\omega )s_{j}^{a},  \label{slf}
\end{equation}
where $G$ is either $A$ or $D$, and the kernel $K^{kj}$ has the same
definition as before (\ref{kdef}), but its indices are running from $-m+1$
to $l-1$. The equation with $j=0$ is now the massive equation.

In the $\mathbf{C}_{n}$\textbf{\ case} equations (\ref{c1}), its analog for $%
\eta ,\widetilde{\eta }$ and (\ref{kca}), can be written as
\begin{equation}
\frac{1}{f}\sum_{b=1}^{n-1}K_{ab}^{C_{n}}\widetilde{s}_{j}^{b}+\frac{1}{f}%
K_{an}^{C_{n}}\widetilde{s}_{j/2}^{n}=-\sum_{k=-2m+1}^{2l-1}K^{jk}\left(
\omega /2\right) s_{k}^{a},\;a\leq n-1  \label{cfa}
\end{equation}
Equations (\ref{c2}), their double for $\eta ,\widetilde{\eta }$, and (\ref
{kcn}) one can write as the following one equation
\begin{equation}
\widetilde{s}_{j}^{n}-\frac{(1/2)}{(1)}\sum_{k=-2m+1}^{2l-1}G^{jk}\widetilde{%
s}_{k}^{n-1}=-\sum_{k=-m+1}^{l-1}K^{jk}\left( \omega \right) s_{k}^{n}.
\label{cfn}
\end{equation}

In the same way the $\mathbf{B}_{n}$\textbf{\ case} equations (\ref{kba1})
and (\ref{b1}),(\ref{b2}) with their $\eta ,\widetilde{\eta }$ analogs give
\begin{equation}
\sum_{b=1}^{n}K_{ab}^{B_{n}}\widetilde{s}_{k}^{b}-\delta _{a,n-1}\frac{(1/2)%
}{(1)}\sum_{j=-2m+1}^{2l-1}G^{kj}\widetilde{s}_{j}^{n}=-%
\sum_{j=1}^{l-1}K^{kj}\left( \omega \right) s_{j}^{a},\;a\leq n-1
\label{bfa}
\end{equation}
and equations (\ref{kbn1}),(\ref{b3}) with $\eta $ partners
\begin{equation}
\widetilde{s}_{k}^{n}-\frac{1}{(1/2)}\widetilde{s}_{k/2}^{n-1}=-%
\sum_{k=-2m+1}^{2l-1}K^{kj}\left( \omega /2\right) s_{j}^{n}.  \label{bfn}
\end{equation}

\bigskip

Doing thermodynamic is standard procedure (see for example \cite{Z},\cite{KM},\cite{M}%
): we should minimize the free energy $F=E-TS$, ($E$ - is energy, $T$ -
temperature, $S$ - entropy), using the derived equations as constraints. The
energy is the energy of massive particles
\[
E=\sum_{a=1}^{n}\int_{-\infty }^{\infty }d\theta s_{0}^{a}(\theta
)m_{a}\cosh \theta,
\]
and the entropy can be calculated from combinatoric of states as particles
and holes in the thermodynamic limit
\[
S=\sum_{a=1}^{n}\sum_{j=-t_{a}m+1}^{t_{a}l-1}\int_{-\infty }^{\infty
}d\theta \left[ \left( s_{j}^{a}+\widetilde{s}_{j}^{a}\right) \ln \left(
s_{j}^{a}+\widetilde{s}_{j}^{a}\right) -s_{j}^{a}\ln s_{j}^{a}-\widetilde{s}%
_{j}^{a}\ln \widetilde{s}_{j}^{a}\right].
\]

We will not repeat the standard free energy minimization procedure. If the
starting constraints equations were written in the form
\[
\sum_{b=1}^{n}Q_{ab}\widetilde{s}_{j}^{b}=-%
\sum_{k=-t_{a}m+1}^{t_{a}l-1}P^{jk}s_{k}^{a},
\]
with some kernels $Q,P$, as it was in eqs. (\ref{slf})--(\ref{bfn}), the
variation leads to the set of equations in the form so called $Y$-system
\begin{equation}
\sum_{b=1}^{n}Q_{ab}L_{(+)j}^{b}=%
\sum_{k=-t_{a}m+1}^{t_{a}l-1}P^{jk}L_{(-)k}^{a},  \label{sys}
\end{equation}
where we defined ''dressed energies'' $\varepsilon _{k}^{a}$, and $L_{(\pm
)k}^{a}$
\[
\frac{\widetilde{s}_{k}^{a}}{s_{k}^{a}}=\exp (\varepsilon
_{k}^{a})=Y_{k}^{a},\;L_{(\pm )k}^{a}=\ln \left( 1+\exp (\pm
\varepsilon _{k}^{a})\right).
\]
The free energy itself \ can be expressed in terms of the stationary values
of functions $Y_{k}^{a}(\pm \infty )$ in the rapidity space. The well known
relation between the central charge of the relativistic theory defined on a
cylinder in the UV limit (radius of the cylinder goes to zero, when
temperature is going to infinity) $F=-\frac{\pi c}{6}T$, gives a possibility
to extract the central charge of the corresponding CFT in the UV limit ($%
\frac{G_{l}\times G_{m}}{G_{l+m}}$ in our case)
\begin{equation}
c=\frac{6}{\pi ^{2}}\sum_{a=1}^{n}\left[ \sum_{j=-t_{a}m+1}^{t_{a}l-1}L%
\left( \frac{1}{1+Y_{j}^{a}}\right) -\sum_{j=-t_{a}m+1,\neq
0}^{t_{a}l-1}L\left( \frac{1}{1+\overline{Y}_{j}^{a}}\right)
\right]. \label{dil}
\end{equation}
Here $L$ is the dilogarithm function, $Y_{j}^{a},\overline{Y}_{j}^{a}$ are
the stationary values of corresponding variables, introduced above, for the
full system, and for the system with removed massive variables $Y_{0}^{a}$.
This result can be considered as the most serious check of the S-matrix, and
the way from S-matrix to the central charge usually has status of conjecture
in the literature, dealing with TBA analysis.

Explicit use of the kernels form for the systems (\ref{slf})--(\ref{bfn})
after their transformation into the thermodynamic ones according to (\ref
{sys}), gives the following $Y$-systems in the rapidity space.

\begin{itemize}
\item  $\mathbf{A}_{n}$
\end{itemize}

\[
R_{j}^{a}(\theta )=\left( 1+Y_{j}^{a+1}(\theta )\right) \left(
1+Y_{j}^{a-1}(\theta )\right).
\]

\begin{itemize}
\item  $\mathbf{D}_{n}$
\end{itemize}

\begin{eqnarray}
R_{j}^{a}(\theta ) &=&\left( 1+Y_{j}^{a+1}(\theta )\right) \left(
1+Y_{j}^{a-1}(\theta )\right) ,\;a\leq n-3, \nonumber\\
R_{j}^{n-2}(\theta ) &=&\left( 1+Y_{j}^{n-3}(\theta )\right) \left(
1+Y_{j}^{n-1}(\theta )\right) \left( 1+Y_{j}^{n}(\theta )\right), \nonumber\\
R_{j}^{n-1}(\theta ) &=&1+Y_{j}^{n-2}(\theta ), \nonumber\\
R_{j}^{n}(\theta ) &=&1+Y_{j}^{n-2}(\theta ). \nonumber
\end{eqnarray}

\begin{itemize}
\item  $\mathbf{C}_{n}$
\end{itemize}

\begin{eqnarray}
R_{j}^{a}(\theta ) &=&\left( 1+Y_{j/2^{\delta _{a,n-1}}}^{a+1}(\theta
)\right) \left( 1+Y_{j}^{a-1}(\theta )\right) ,\;a\leq n-1, \nonumber\\
R_{j}^{n}(\theta ) &=&\left( 1+Y_{2j}^{n-1}(\theta +i/2)\right)
\left( 1+Y_{2j}^{n-1}(\theta -i/2)\right) \left(
1+Y_{2j+1}^{n-1}(\theta )\right) \left( 1+Y_{2j-1}^{n-1}(\theta
)\right). \nonumber
\end{eqnarray}

\begin{itemize}
\item  $\mathbf{B}_{n}$
\end{itemize}

\begin{eqnarray}
R_{j}^{a}(\theta ) &=&\left( 1+Y_{j}^{a+1}(\theta )\right) \left(
1+Y_{j}^{a-1}(\theta )\right) \ast \nonumber\\
&&\left[ \left( 1+Y_{2j}^{n}(\theta +i/2)\right) \left( 1+Y_{2j}^{n}(\theta
-i/2)\right) \left( 1+Y_{2j+1}^{n}(\theta )\right) \left(
1+Y_{2j-1}^{n}(\theta )\right) \right] ^{\delta _{a,n-1}}, \nonumber\\
R_{j}^{n}(\theta ) &=&\left( 1+Y_{j/2}^{n-1}(\theta )\right).
\nonumber
\end{eqnarray}
Here
\[
R_{j}^{a}(\theta )=\left( 1+\frac{1}{Y_{j+1}^{a}(\theta )}\right) \left( 1+%
\frac{1}{Y_{j-1}^{a}(\theta )}\right) Y_{j}^{a}(\theta
+i/t_{a})Y_{j}^{a}(\theta -i/t_{a}).
\]
One can see that left hand side $R_{j}^{a}$ of these equations contains
different $j$ indexes and the same index $a$, whilst their right hand side
contains different $a$ indexes. Traditionally these TBA equations one
represents schematically as TBA diagram (see fig. 1,2). Their nodes
correspond to each $Y_{j}^{a}$. Those $j$'s which appear in the $R_{j}^{a}$
along with $j$ itself, are depicted by vertical lines connecting $(a,j)$
node with others. Horizontal and other lines represent other indices then $a$
and $j$ appearing in the r.h.s. of $(a,j)$ equation. On these figures
massive nodes are depicted as bold ones, and magnonic -- as grey ones. The
purely magnonic equations are represented by the same TBA diagrams with
removed massive nodes.

\begin{figure}[tbp]
\epsfysize=1.5 true in \hskip 20 true pt \epsfbox{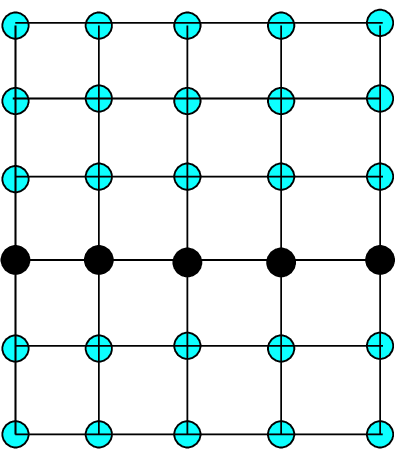} \hskip 30 true pt %
\epsfysize=1.5 true in \hskip 20 true pt \epsfbox{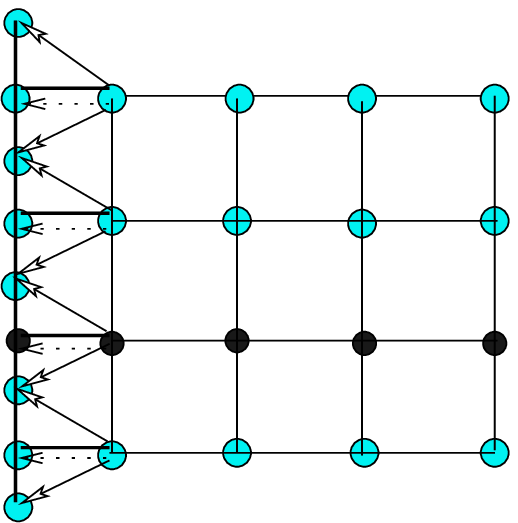} 
\caption{TBA diagrams for the cases \textbf{\ a} (left) $G=A_5,l=4,m=3$,
\textbf{\ b} (right) $G=B_5,l=3,m=2$}
\label{F1}
\end{figure}
\begin{figure}[tbp]
\epsfysize=1.5 true in \hskip 20 true pt \epsfbox{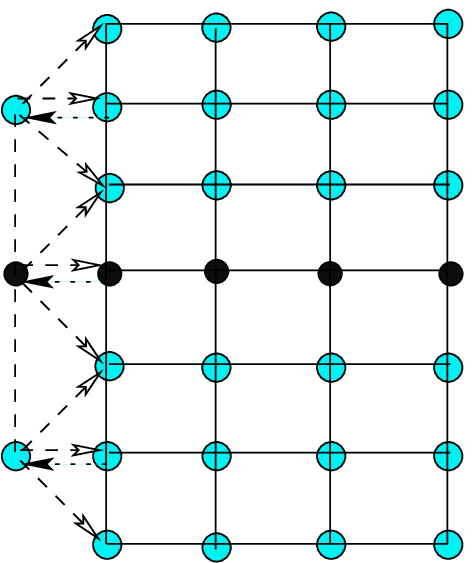} \hskip 30 true pt %
\epsfysize=1.5 true in \hskip 20 true pt \epsfbox{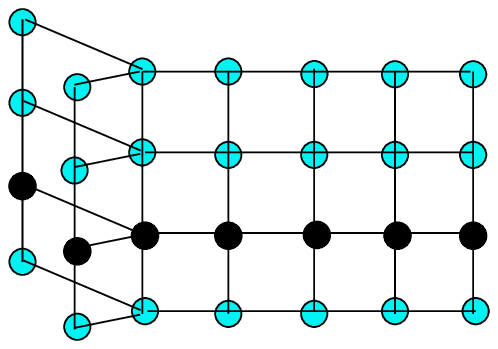} 
\caption{TBA diagrams for the cases \textbf{\ a} (left) $G=C_5,l=2,m=2$,
\textbf{\ b} (right) $G=D_7,l=3,m=2$}
\label{F2}
\end{figure}

As one can check, the above $Y$-systems exactly reproduce those which were
considered in \cite{KN}, if one changes $Y_{j}^{a}\rightarrow \left(
Y_{j}^{a}\right) ^{-1}$. In \cite{KN} such $Y$-systems were related to
quantum affine Lie algebras $U_{q}(G_{n}^{(1)})$ and their representations.
The mathematical meaning standing behind functional $Y$-systems and their
relation to representation theory of quantum affine Lie algebras, their
character relations and transfer matrix functional relations, is deep and
beautiful issue, requiring further investigation. Here we will cite the main
result of \cite{KN}: the dilogarithm sum rules like (\ref{dil})
corresponding to the ground state of the $Y$-systems we obtained (we omit
some technical details and present it in a simplified form relevant for our
case):
\begin{equation}
\frac{6}{\pi ^{2}}\sum_{a=1}^{n}\sum_{j=1}^{t_{a}l-1}L\left( \frac{X_{j}^{a}%
}{1+X_{j}^{a}}\right) =\frac{l\dim G_{n}}{l+g}-n,  \label{kn}
\end{equation}
where $X_{j}^{a}$ satisfy the same functional relations as our $\left(
Y_{j}^{a}\right) ^{-1}$. Using the relation $L(t)=1-L(1-t)$ and substituting
(\ref{kn}) into (\ref{dil}), one can see that the latter reproduce the
correct central charge
\[
c=\dim G_{n}\left(
\frac{l}{l+g}+\frac{m}{m+g}-\frac{l+m}{l+m+g}\right)
\]
for all the coset models considered here.

This completes the derivation of TBA $Y$-systems from the S-matrices of
relativistic integrable models, and their check by comparison of their
ground state thermodynamic with the central charge of their UV limit CFTs.
This is one of the most convenient checks of \ the S-matrix correctness.

\section{\protect\bigskip Full S-matrix}

As we have seen in the previous section, the consistent TBA, which
reproduces the correct central charge, dictates unique form of the S-matrix
crossing-unitarizing prefactor together with CDD facor, for any pair of two
fundamental representations for any algebra $G$. Here we will show that in
the cases where the full S-matrix is known for any pair of fundamental
representations, i.e. in multiplicity free cases $A_{n},C_{n}$, this
prefactor can be calculated from the S-matrix. In the $A_{n}$ case this
calculation was done in \cite{H}. The most interesting information one can
extract from the above TBA analysis, is the prefactor for $B_{n},D_{n}$
cases. Although \ the S-matrix $S_{ab}$ is not know in this cases for any $%
a,b$, it is known, for example, for spinor-spinor S-matrices: in $B_{n}$
case for $a=b=n$, or in $D_{n}$ case -- $a=b=n$ and $a=n,b=n-1$, since then
spectral decomposition is multiplicity free. Corresponding rational
S-matrices were constructed for PCM in the seminal work \cite{ORW}, but we,
unfortunately, don't know any published trigonometric generalization of
their result. It would be interesting to work out corresponding
trigonometric S-matrices, especially in their RSOS form relevant in our case.

We recall that our starting S-matrix has the form
\[
S_{ab}(\theta )=X_{ab}(\theta )S_{ab}^{(l)}(\theta )\otimes
S_{ab}^{(m)}(\theta ).
\]
The total crossing-unitarizing prefactor $Y_{ab}$ is the product of CDD
factor $X_{ab}$ and corresponding crossing-unitarizing prefactors $\sigma
_{ab}^{(l)},\sigma _{ab}^{(m)}$ for each $S_{ab}^{(l)}(\theta )$ and $%
S_{ab}^{(m)}$. As was shown above, the correct TBA derivation, confirming
the proper central charge in all the considered cases, was based on the
assumption for the crossing-unitaring prefactor form of the full S-matrix (%
\ref{asum}) with the $\varphi $ having the universal form for all the
algebras (\ref{phi}). First, we start with a proof of this statement for $%
A_{n}$ and $C_{n}$ cases, when the full S-matrix $S_{ab}^{(l,m)}(\theta )$
is known for any pair $a,b$, since being free of multiplicities, the
spectral decomposition, and crossing-unitarizing including prefactors $%
\sigma _{ab}^{(l,m)}$, can be calculated explicitly.

\subsection{Vector-vector S-matrices and their fusion}

Recall the general RSOS structure of the S-matrix $S_{ab}^{(l)}(\theta )$
for any algebra $G_{n}$. In general one gets $S_{ab}^{(l)}(\theta )$ as a
result of fusion procedure starting from the fundamental S-matrix. By
fundamental one means the S-matrix for fundamental representations, from
which all other representations will appear after decomposition of tensor
product of reps into irreducible ones. This representations are usually
called the defining representations. The defining reps for $A_{n}$ and $%
C_{n} $ algebras are their vector representations, while for $B_{n}$ and $%
D_{n}$ -- their spinor ones. In the last case all other fundamental
representations can be obtained as a tensor product of spinor ones. But,
obtaining the vector representation in the tensor product of spinors, one
can get from the vector representation all the others by the same fusion
procedure, except for the spinor ones. We don't know the explicit form of
the spinor-spinor trigonometric RSOS S-matrices for $B_{n}$ and $D_{n}$
cases (one can guess they look quite complicated), although they are, as we
said, the fundamental S-matrices in these cases. Instead, we will describe
vector-vector S-matrices for all the cases. The structure of RSOS S-matrix
of the vector-vector representation is well known and described in the
literature (see, for example, \cite{JMO},\cite{BR}). We cite it here for
completeness and reader convenience, following \cite{HolS}. It is defined as
scattering process of two kinks
\[
K_{ac}(\theta _{1})+K_{cd}(\theta _{2})\rightarrow K_{ab}(\theta
_{2})+K_{bd}(\theta _{1})
\]
connecting different vacua $a,b,c,d\in \Lambda ^{\ast }$ of the theory from
the weight lattice $\Lambda ^{\ast }$ of the algebra $G_{n}$. Weights of its
vector representation are
\begin{eqnarray}
\Sigma &=&\{e_{1}-e_{0},...,e_{n+1}-e_{0}\},\;e_{0}=\frac{%
\sum_{i=1}^{n+1}e_{i}}{n+1},\;A_{n}, \nonumber\\
\Sigma &=&\{0,\pm e_{1},...,\pm e_{n}\},\;B_{n}, \nonumber\\
\Sigma &=&\{\pm e_{1},...,\pm e_{n}\},\;C_{n},D_{n}, \nonumber
\end{eqnarray}
where ${e_{i}}$ is some orthonormal basis. Up to a prefactor scalar function
$Y $\ this kink-kink S-matrix is proportional to the Boltsmann weights (BW) $%
W$ of statistical lattice models with corresponding symmetry, constructed as
solutions of Yang-Baxter interaction round the face equations \cite{JMO}:
\begin{equation}
S_{11}^{(l)}(\theta _{2}-\theta _{1})=Y(u)W\left( \left.
\begin{array}{cc}
a & b \\
c & d
\end{array}
\right| f(u)\right) \left( \frac{G_{a}G_{d}}{G_{b}G_{c}}\right) ^{f(u)/2},
\label{smw}
\end{equation}
where $u=(\theta _{2}-\theta _{1})/i\pi ,f(u)=cu$ for some constant $c$, $%
c-a,d-c,b-a,d-b\in \Sigma $, and $G_{a}$ will be defined below. The set of
non zero BW for vector-vector representation looks as follows \cite{JMO}:
\begin{eqnarray}
W\left( \left.
\begin{array}{cc}
a & a+\mu \\
a+\mu & a+2\mu
\end{array}
\right| u\right) &=&\frac{\sin (\omega -\lambda u)}{\sin (\omega )}
\label{bwa} \\
W\left( \left.
\begin{array}{cc}
a & a+\mu \\
a+\mu & a+\mu +\nu
\end{array}
\right| u\right) &=&\frac{\sin (a_{\mu \nu }+\lambda u)}{\sin (a_{\mu \nu })}
\nonumber\\
W\left( \left.
\begin{array}{cc}
a & a+\nu \\
a+\mu & a+\mu +\nu
\end{array}
\right| u\right) &=&\frac{\sin (\lambda u)}{\sin (\omega )}\left( \frac{\sin
(a_{\mu \nu }+\omega )\sin (a_{\mu \nu }-\omega )}{\sin ^{2}(a_{\mu \nu })}%
\right) ^{1/2} \nonumber
\end{eqnarray}
($\mu \neq \nu $) for $A_{n}$ case, and
\begin{eqnarray}
W\left( \left.
\begin{array}{cc}
a & a+\mu \\
a+\mu & a+2\mu
\end{array}
\right| u\right) &=&\frac{\sin (\lambda -\lambda u)\sin (\omega -\lambda u)}{%
\sin (\omega )\sin (\lambda )},\;\mu \neq 0  \nonumber \\
W\left( \left.
\begin{array}{cc}
a & a+\mu \\
a+\mu & a+\mu +\nu
\end{array}
\right| u\right) &=&\frac{\sin (\lambda -\lambda u)\sin (a_{\mu
\nu }+\lambda u)}{\sin (\lambda )\sin (a_{\mu \nu })},\;\mu \neq
\pm \nu  \nonumber
\\
W\left( \left.
\begin{array}{cc}
a & a+\nu \\
a+\mu & a+\mu +\nu
\end{array}
\right| u\right) &=&\frac{\sin (\lambda -\lambda u)\sin (\lambda u)}{\sin
(\omega )\sin (\lambda )}\times  \nonumber \\
\times \left( \frac{\sin (a_{\mu \nu }+\omega )\sin (a_{\mu \nu }-\omega )}{%
\sin ^{2}(a_{\mu \nu })}\right) ^{1/2}\ \mu &\neq &\pm \nu  \label{bw}
\end{eqnarray}
\begin{eqnarray}
W\left( \left.
\begin{array}{cc}
a & a+\nu \\
a+\mu & a
\end{array}
\right| u\right) &=&\frac{\sin (\lambda u)\sin (a_{\mu -\nu }+\omega
-\lambda +\lambda u)}{\sin (\lambda )\sin (a_{\mu -\nu }+\omega )}\left(
G_{a_{\mu }}G_{a_{\nu }}\right) ^{1/2}+ \nonumber\\
+\delta _{\mu \nu }\frac{[\lambda -\lambda u][a_{\mu -\nu }+\omega +\lambda
u]}{[\lambda ][a_{\mu -\nu }+\omega ]},\;\mu &\neq &0 \nonumber\\
W\left( \left.
\begin{array}{cc}
a & a \\
a & a
\end{array}
\right| u\right) &=&\frac{\sin (\lambda +\lambda u)\sin (2\lambda -\lambda u)%
}{\sin (\lambda )\sin (2\lambda )}-\frac{\sin (\lambda -\lambda
u)\sin (\lambda u)}{\sin (\lambda )\sin (2\lambda )}J_{a}
\nonumber
\end{eqnarray}
for $B_{n},C_{n}$ and $D_{n}$, where $\mu ,\nu \in \Sigma ,a_{\mu }=\omega
(a+\rho )\cdot \mu $, ($\rho $ is the sum of the fundamental weights of the
algebra), $a_{0}=-\omega /2,a_{\mu \nu }=a_{\mu }-a_{\nu },a_{\mu -\nu
}=a_{\mu }+a_{\nu }$, and the constants $\omega ,\lambda $ are defined as
\[
\omega =\frac{\pi }{t(g+l)},\lambda =\frac{tg\omega }{2}.
\]
Parameter $t=1$ in $A_{n},B_{n},D_{n}$ cases and $t=2$ in $C_{n}$ case. We
also used
\begin{eqnarray}
G_{a_{\mu }} &=&\sigma \frac{s(a_{\mu }+\omega )}{s(a_{\mu })}\prod_{\kappa
\neq \pm \mu ,0}\frac{\sin (a_{\mu \kappa }+\omega )}{\sin (a_{\mu \kappa })}%
,\;\mu \neq 0,\;G_{a_{0}}=1, \nonumber\\
J_{a} &=&\sum_{\kappa \neq 0}\frac{\sin (a_{\kappa }+\omega /2-2\lambda )}{%
\sin (a_{\kappa }+\omega /2)}G_{a_{\kappa }}, \nonumber
\end{eqnarray}
where $\sigma =-1$ in the $C_{n}$ case and $\sigma =1$ in the $B_{n},D_{n}$
cases. The function $s(x)=\sin (tx)$ in the $B_{n},C_{n}$ cases, and $s(x)=1$
for $D_{n}$. The quantities $G_{a}$ used in (\ref{smw}) are related to $%
G_{a_{\mu }}$ by $G_{a_{\mu }}=G_{a+\mu }/G_{a}$ and can be written as
\[
G_{a}=\varepsilon (a)\prod_{i=1}^{n(+1)}s(a_{i})\prod_{1\leq
i<j\leq n(+1)}\sin (a_{i}-a_{j})\sin (a_{i}+a_{j}),
\]
where $a=\sum_{i=1}^{n(+1)}a_{i}e_{i}$ defines $a_{i}$, and $\varepsilon (a)$
is a sign factor chosen so that $\varepsilon (a+\mu )/\varepsilon (a)=\sigma
$.

The models are called restricted since only the dominant weights
\begin{equation}
a\cdot \theta \leq l  \label{res}
\end{equation}
are allowed, where $\theta $ is the highest root of the algebra, and $l$ is
the level.

The BW listed above satisfy a set of conditions important for the S-matrix
construction:

\begin{itemize}
\item  unitarity
\begin{equation}
\sum_{e}W\left( \left.
\begin{array}{cc}
a & e \\
c & d
\end{array}
\right| u\right) W\left( \left.
\begin{array}{cc}
a & b \\
e & d
\end{array}
\right| -u\right) =\rho (u)\delta _{bc},  \label{u}
\end{equation}
where in $A_{n}$ case
\[
\rho (u)=\frac{\sin (\omega -\lambda u)\sin (\omega +\lambda
u)}{\sin ^{2}(\omega )},
\]
\end{itemize}

and in $B_{n},C_{n},D_{n}$ cases
\begin{equation}
\rho (u)=\frac{\sin (\omega -\lambda u)\sin (\omega +\lambda u)\sin (\lambda
-\lambda u)\sin (\lambda +\lambda u)}{\sin ^{2}(\omega )\sin ^{2}(\lambda )},
\label{ro}
\end{equation}

\begin{itemize}
\item  crossing symmetry ($B_{n},C_{n},D_{n}$ cases):
\begin{equation}
W\left( \left.
\begin{array}{cc}
a & b \\
c & d
\end{array}
\right| u\right) =W\left( \left.
\begin{array}{cc}
c & a \\
d & b
\end{array}
\right| 1-u\right) \left( \frac{G_{b}G_{c}}{G_{a}G_{d}}\right) ^{1/2}.
\label{cr}
\end{equation}

\item  crossing-unitarity relation ($A_{n}$ case)
\begin{equation}
\sum_{e}W\left( \left.
\begin{array}{cc}
a & e \\
c & d
\end{array}
\right| 1+u\right) W\left( \left.
\begin{array}{cc}
a & b \\
e & d
\end{array}
\right| 1-u\right) =\delta _{bc}\frac{\sin (\lambda -\lambda u)\sin (\lambda
+\lambda u)}{\sin ^{2}(\omega )}.  \label{uc}
\end{equation}
\end{itemize}

The case $A_{n}$ is different, since vector representation is not conjugate
to itself.

Requirements of unitarity and crossing for the S-matrix (\ref{smw}), using (%
\ref{u}) and (\ref{uc}) in the $A_{n}$ case, lead to the following
functional constraint on the function $Y(u)$, to which we will add two
indices $Y_{n,l}(u)$, explicitly emphasizing its dependence on rank $n$ and
level $l$:
\begin{eqnarray}
Y_{n,l}(u)Y_{n,l}(-u) &=&\frac{\sin ^{2}(\omega )}{\sin (\omega +\lambda
u)\sin (\omega -\lambda u)}, \nonumber\\
Y_{n,l}(1+u)Y_{n,l}(1-u) &=&\frac{\sin ^{2}(\omega )}{\sin
(\lambda +\lambda u)\sin (\lambda -\lambda u)}. \nonumber
\end{eqnarray}
The minimal solution of this system of functional relations (up to so called
CDD umbiguities), which has no poles on the physical strip $0<u<1$, was
found in \cite{DVF}:
\begin{eqnarray}
Y_{n,l}(u) &=&\exp \left\{ 2\int_{0}^{\infty }\frac{dx}{x}\frac{[\frac{n+1}{2%
}u]}{[l+n+1][n+1]}\times \right.  \label{ynl} \\
&&\left. \times \left\{ (l)\left( \frac{n+1}{2}u\right) -(n-1+l)\left( \frac{%
u(n+1)}{2}-n-1\right) \right\} \right\}.  \nonumber
\end{eqnarray}
We recall that we use the short notations $\left[ z\right] =\sinh
(xz),\left( z\right) =\cosh (xz)$.

In the same way the functional relations on $Y$ in the $B_{n},C_{n},D_{n}$
cases, using \ (\ref{u}),(\ref{cr}), look like
\begin{eqnarray}
Y(u) &=&Y(1-u) \nonumber\\
Y(u)Y(-u) &=&\rho ^{-1}(u), \nonumber
\end{eqnarray}
where $\rho $ is defined in (\ref{ro}). The minimal solution of this system
can be written in terms of \ $Y_{n,l}$ (\ref{ynl})
\begin{equation}
Y(u)=Y_{tg,tl}(u)Y_{tg,tl}(1-u)\frac{\sin \lambda }{\sin \omega }.
\label{ybcd}
\end{equation}

The S-matrices described here can be written in the spectral decomposed form
with projectors onto irreducible representations appearing in the tensor
product of two vector representations. (These projectors should be written
in the IRF form). As was pointed out, the constructed vector-vector S-matrix
is pole free and has no bound states. The bound states are produced by
insertion of CDD factors, which have poles at the rapidity value
corresponding to the desired projector in the spectral decomposition. It is
well known that the closed and self consistent bootstrap procedure requires
the pole in $S_{11}$ S-matrix in the channel corresponding to the second
fundamental weight. Continuation of the bootstrap reproduces massive bound
state multiplets corresponding to all the fundamental representations of the
algebra. This scenario was proved to be correct in the cases where the
spectral decomposition of $S_{ab}$ is known for any $a,b$, and was
conjectured to be correct in all the cases. One of the most effective
methods of the construction of S-matrix spectral decomposition is the tensor
product graph method (TPG). But unfortunately it works only when the
spectral decomposition is multiplicity free. This is the situation in the $%
A_{n},C_{n}$ cases where the tensor product of two representations with
fundamental highest weights are multiplicity free.

We briefly recall here the concept of TPG. First of all, there is
requirement for the representations, for which we build TPG, to be \textit{%
affinizable} (for details see \cite{MK},\cite{ZGB}). All the fundamental
representations of $A_{n}$ and $C_{n}$ are affinizable. Unfortunately, it is
not the case for $B_{n}$ and $D_{n}$ cases. TPG is a graph which is
constructed by letting the irreducible components $\lambda $ and $\sigma $
of tensor product $\lambda _{a}\otimes \lambda _{b}$ be the nodes of the
graph joined by \ a link if $\lambda $ and $\sigma $ have of opposite parity
and $\sigma \subset adj\otimes \lambda $. The parity of $\lambda $ is
defined to be $\pm 1$ according to whether $\lambda $ appears symmetrically
or anti-symmetrically in the tensor product in the limit $q\rightarrow -1$.
The concept of TPG works well when TPG is a tree and there are no
multiplicities in the tensor product of two representations. In the Tables
1,2 one can find the TPG for any pair of fundamental weights of algebras $%
A_{n}$ and $C_{n}$.

\bigskip \bigskip

\begin{tabular}{|llllll|}
\hline
$\lambda _{a}+\lambda _{b}$ & $\rightarrow $ & $\lambda _{a+1}+\lambda
_{b-1} $ & $...$ & $\rightarrow $ & $\lambda _{a+\min (n+1-a,b)}+\lambda
_{b-\min (n+1-a,b)}$ \\ \hline
\end{tabular}

\bigskip

Table 1. Tensor product graph for two fundamental representations $a$ and $b$
of $A_{n}$ ($a\geq b$).

\bigskip \bigskip
\begin{tabular}{|llllllll|}
\hline
$\lambda _{a}+\lambda _{b}$ & $\rightarrow $ & $\lambda _{a+1}+\lambda _{b-1}
$ & $\rightarrow $ & $...$ & $\lambda _{a+b-1}+\lambda _{1}$ & $\rightarrow $
& $\lambda _{a+b}$ \\
$\downarrow $ &  & $\downarrow $ &  &  & $\downarrow $ &  &  \\
$\lambda _{a-1}+\lambda _{b-1}$ & $\rightarrow $ & $\lambda _{a}+\lambda
_{b-2}$ & $\rightarrow $ & $...$ & $\lambda _{a+b-2}$ &  &  \\
$\downarrow $ &  & $\downarrow $ &  &  &  &  &  \\
$\vdots $ &  & $\vdots $ &  &  &  &  &  \\
$\lambda _{a-b+1}+\lambda _{1}$ & $\rightarrow $ & $\lambda _{a-b+2}$ &  &
&  &  &  \\
$\downarrow $ &  &  &  &  &  &  &  \\
$\lambda _{a-b}$ &  &  &  &  &  &  &  \\ \hline
\end{tabular}

\bigskip

Table 2. Tensor product graph for two fundamental representations $a$ and $b$
of $C_{n}$ ($a\geq b$). For $a+b>n$ the graph truncates at the ($n-a+1$)th
column.

\bigskip

Given a TPG, the spectral decomposition of the R-matrix has the form
\[
R_{ab}(x)=\sum_{\mu }\rho _{\mu }(x)P_{\mu },
\]
where the sum is over the irreps appearing in the tensor product
decomposition, i.e. over the nodes of TPG, $x$ is multiplicative spectral
parameter, and the main rule dictated by TPG says: if there is an arrow from
$\mu $ to $\nu $ on the TPG then the coefficients $\rho _{\mu }(x)$ and $%
\rho _{\nu }(x)$ satisfy
\[
\frac{\rho _{\mu }(x)}{\rho _{\nu }(x)}=\frac{xq^{I(\mu
)/2}-x^{-1}q^{I(\nu )/2}}{x^{-1}q^{I(\mu )/2}-xq^{I(\nu )/2}},
\]
where
\[
x=e^{i\lambda u},\;q=-e^{-i\omega },
\]
and the second Casimir $I(\mu )=(\mu +2\rho )\cdot \mu $.

Using these TPG method one can construct the minimal S-matrix for scattering
of any two fundamental multiplets of $A_{n}$ and $C_{n}$ \cite{HolS}. It has
the form:
\[
S_{ab}^{(l)}(u)=\sigma _{ab}^{(l)}(u)R_{ab}(u),
\]
where
\begin{equation}
\sigma _{ab}^{(l)}(u)=Z_{ab}(u)\prod_{j=1}^{a}\prod_{k=1}^{b}Y\left( u+\frac{%
2j+2k-a-b-2}{tg}\right)  \label{sig}
\end{equation}
and $Y$ is defined by (\ref{ynl}) in $A_{n}$ case, and by (\ref{ybcd}) -- in
$C_{n}$ case. $Z_{ab}$ coming from the R-matrix, is defined below. The
R-matrices have the following form ($b\geq a$)

\begin{itemize}
\item  $\mathbf{A}_{n}$\textbf{\ case}
\[
R_{ab}(u)=\sum_{k=0}^{\min (n+1-b,a)}(-1)^{k+1}\rho
_{ab}^{k}(u)P_{\lambda _{b+k}+\lambda _{a-k}}
\]
\end{itemize}

(by definition $\lambda _{n+1}=\lambda _{0}=0$) with
\[
\rho _{ab}^{k}(u)=\prod_{p=1}^{k}\frac{\{2p+b-a\}}{\{-2p-b+a\}},
\]

and $Z_{ab}$ in this case
\begin{equation}
Z_{ab}(u)=\prod_{j=1}^{a}\prod_{k=1}^{b-1}\{2j+2k-a-b\}\prod_{p=1}^{a}%
\{-2p-b+a\}.  \label{za}
\end{equation}

Here and below we use a new notation
\[
\{x\}=\frac{\sin (\omega x/2+\lambda u)}{\sin (\omega )}.
\]

\begin{itemize}
\item  $\mathbf{C}_{n}$\textbf{\ case}
\begin{eqnarray}
R_{ab}(u) &=&\sum_{j=0}^{\min (n-b,a)}\sum_{k=0}^{a-j}(-1)^{k+j}\rho
_{ab}^{jk}(u)P_{\lambda _{b+j-k}+\lambda _{a-j-k}}, \nonumber\\
\rho _{ab}^{jk}(u) &=&\prod_{p=1}^{j}\frac{\{2p+b-a\}}{\{-2p-b+a\}}%
\prod_{q=1}^{k}\frac{\{2(n+1)+2q-b-a\}}{\{-2(n+1)-2q+b+a\}},
\nonumber
\end{eqnarray}
\end{itemize}

and $Z_{ab}$ in this case
\begin{eqnarray}
Z_{ab}(u) &=&\left( \frac{\sin \lambda }{\sin \omega }\right)
^{ab}\prod_{j=1}^{a}\prod_{k=1}^{b-1}\{2j+2k-a-b\}\{a+b-2(n+1)-2j-2k\}
\nonumber \\
&&\times \prod_{p=1}^{a}\{-2p-b+a\}\{-2(n+1)-2p+a+b\}.  \label{zc}
\end{eqnarray}

\begin{itemize}
\item  $\mathbf{B}_{n}$\textbf{\ and }$\mathbf{D}_{n}$\textbf{\ cases}
\end{itemize}

In these cases affinizable among fundamental weights representations are
only vector and spinor ones, and their tensor products are multiplicity
free. We will consider here only S-matrix for two vector representations. It
has the following spectral decomposition \cite{ORW}\cite{HolS}:
\[
S_{11}^{(l)}(u)=\{-2\}\{-g\}Y_{g,l}(u)Y_{g,l}(1-u)\left[ P_{2\lambda _{1}}-%
\frac{\{2\}}{\{-2\}}P_{\lambda _{2}}+\frac{\{2\}\{g\}}{\{-2\}\{-g\}}P_{0}%
\right].
\]

\subsection{CDD factors}

The S-matrices presented above are crossing symmetric, unitary and minimal
in the sense that they have no poles in the physical strip. As we said, in
order to make bootstrap working, one should multiply these S-matrices by a
set of properly chosen CDD factors. They appear as a result of fusion of the
CDD factors for vector-vector S-matrices. It is useful to introduce the
following notation for universal description of CDD factors
\[
\overline{X}(a)=\frac{\sin \left( \frac{\pi }{2}\left(
u+\frac{a}{tg}\right) \right) }{\sin \left( \frac{\pi }{2}\left(
u-\frac{a}{tg}\right) \right) }.
\]
(We recall that $t=1$ in $A,B,D$ cases and $t=2$ in $C$ case.)

\begin{itemize}
\item  $\mathbf{A}_{n}$\textbf{\ case}
\end{itemize}

Vector-vector S-matrix CDD factor is just $\overline{X}(2)$ and its fusion
leads to the following CDD factor for $a,b$ scattering
\[
X_{ab}(u)=\prod_{j=|a-b|+1,\;step\;2}^{a+b-1}\overline{X}(j+1)%
\overline{X}(j-1).
\]
One can see that there is a pole in $X_{ab}(u)$ at $u=\frac{a+b}{n+1}$ if $%
a+b<n+1$, or at $u=2-\frac{a+b}{n+1}$ if $a+b>n+1$. They correspond to
particles $a+b$ or $a+b-n-1$ respectively in the direct channel. There is
also a pole at $u=\frac{|a-b|}{n+1}$ corresponding to the particle $|a-b|$
in the cross channel. We will not discuss here the double poles, this
discussion one can find in \cite{HolS}.

\begin{itemize}
\item  $\mathbf{C}_{n}$\textbf{\ case}
\end{itemize}

Here the vector-vector CDD factor has the form $X_{11}=\overline{X}(2)%
\overline{X}(2g-2)$ and its fusion gives
\begin{equation}
X_{ab}(u)=\prod_{j=|a-b|+1,\;step\;2}^{a+b-1}\overline{X}(j+1)%
\overline{X}(j-1)\overline{X}(2(n+1)-j+1)\overline{X}(2(n+1)-j-1).
\label{cddc}
\end{equation}
As one can see from (\ref{cddc}), the pole structure here is more
complicated -- it exhibits poles up to the 4-th order. Discussion of the
S-matrix analytic structure in this case one can find in \cite{HolS}.

\begin{itemize}
\item  $\mathbf{B}_{n}$\textbf{\ and }$\mathbf{D}_{n}$\textbf{\ cases. }
\end{itemize}

CDD factors for the vector-vector S-matrices in these cases has the same
form $X_{11}=\overline{X}(2)\overline{X}(2g-2)$.

\subsection{Prefactor exponentialization}

Now we have to transform both $X_{ab}$ and $\sigma _{ab}^{(l)}$ to the
exponential form in order to calculate the quantity $Y_{ab}=$ $\frac{1}{2\pi
i}\frac{d}{d\theta }\left( X_{ab}\sigma _{ab}^{(l)}\sigma _{ab}^{(m)}\right)
$ used in the TBA calculations.One can use for that the following identity
\[
\sin (\pi a)=\exp \left( -\int_{0}^{\infty
}\frac{dx}{x}\frac{2\sinh ^{2}\left( x\left( a-1/2\right) \right)
}{\sinh x}\right)
\]
valid for $0<a<1$, and also
\[
\frac{\sinh (\lambda (\theta +i\pi \alpha ))}{\sinh (\lambda (\theta -i\pi
\alpha ))}=\exp \left\{ -2\int_{0}^{\infty }\frac{dx}{x}\sinh (i\theta x)%
\frac{\sinh \left( \frac{\pi x}{2}(\frac{1}{\lambda }-2\alpha )\right) }{%
\sinh \left( \frac{\pi x}{2\lambda }\right) }\right\}
\]
valid for $0<\alpha <\frac{1}{\lambda }$. Straightforward but long
calculations, using (\ref{sig}),(\ref{za}),(\ref{zc}),(\ref{ynl}),(\ref{ybcd}%
) lead to the following exponential representations valid in both $A_{n}$
and $C_{n}$ cases:

\begin{eqnarray}
X_{ab}(u) &=&\exp \left( 2\int_{0}^{\infty }\frac{dx}{x}e^{-ux}\left( \delta
_{ab}-2\coth x\widetilde{A}_{ab}^{G}(x)\right) \right) ,  \label{cdd} \\
\sigma _{ab}^{(l)}(u) &=&\exp \left( 2\int_{0}^{\infty }\frac{dx}{x}e^{-ux}%
\widetilde{A}_{ab}^{G}(x)\frac{a_{1}^{(g+l)}(x)}{[1]}\right) ,  \label{sigma}
\end{eqnarray}
where the kernels $\widetilde{A}_{ab}^{G}$ are defined in Appendix and
already appeared in the TBA calculations. It turns out that in $B_{n}$ and $%
D_{n}$ cases the vector-vector S-matrices fit the same general formulas (\ref
{cdd}),(\ref{sigma}). One can see after some simple algebra, that these
expressions give a universal answer for the quantity valid in all the cases
described above
\[
\frac{1}{2\pi i}\frac{d}{d\theta }\ln \left( X_{ab}\sigma
_{ab}^{(l)}\sigma _{ab}^{(m)}\right) =\delta _{ab}\delta (\theta
)-A_{ab}^{G}\ast \left( A_{g+l,g+l}^{A_{2g+l+m-1}}\right) ^{-1},
\]
where $\ast $ means convolution. This coincides with the assumption made
about the form of $Y_{ab}$ in the TBA calculations of the previous section.

\subsection{Special cases of S-matrices}

As we mentioned in the introduction, the form of S-matrix (\ref{sm})
includes in it as subclasses S-matrices for other important two dimensional
integrable models with Yangian symmetries. One of them are PCM ($%
l,m\rightarrow \infty $). They are well studied, well defined for any $%
G=A,B,C,D$ and their S-matrices are self consistent from the bootstrap point
of view (see \cite{ORW}). In particular, there are no double poles
unexplainable by Coleman-Thun mechanism. The situation is more subtle with
GN models ($l\rightarrow \infty ,m=1$). Here the S-matrix conjecture
\begin{equation}
S_{ab}=X_{ab}S_{ab}^{(\infty )},  \label{gn0}
\end{equation}
wich was naively expected to be correct in all the cases $G=A,B,D$ (see the
footnote in the introduction about the $C_{n}$ case), was found to suffer
from the ''bootstrap violation'' \cite{ORW} in the $B_{n}$ case, while was
shown to be correct in the $A$ and $D$ cases. Only recently the $B_{n}$ GN
S-matrix was ''corrected'' \cite{FS} using elegant physical arguments about
the symmetry of the Lagrangian. The presence of additional current (compared
with $D_{n}$ case) makes the symmetry different, leading to additional RSOS
like vacuum degeneracy with additional kink structure. The answer for the
S-matrix was shown to be
\begin{equation}
S_{ab}=X_{ab}S_{ab}^{(\infty )}\otimes \widetilde{S}_{ab},  \label{gn}
\end{equation}
where $\widetilde{S}_{ab}=1$ for any pair ($a,b)$ except for ($n,n)$, when $%
\widetilde{S}_{nn}=S_{TCI}(\lambda \theta )$ - the RSOS S-matrix of the
tricritical Ising model with rescaled rapidity. From the Lie algebraic point
of view it is just $A_{1}$ level 2 RSOS model described in the formulas (\ref
{bwa}). There are some identities and specific features of low level RSOS
models. In particular, one can show that $B_{n}$ level 1 RSOS model is
identical (up to a rescaling of the spectral parameter $u$) to $A_{1}$ level
2 RSOS model. It is interesting to note that there are the same identities
on the level of affine Lie algebras themselves, which were pointed out in
the context of generalized parafermions in \cite{G}. In this sense the
S-matrix constructed in \cite{FS} naturally fits the general form of our
S-matrix (\ref{sm}). The identity between (\ref{gn0}) and (\ref{sm}) in $%
A_{n},D_{n}$ cases follows from the fact that RSOS level 1 $A_{n}$ and $%
D_{n} $ S-matrices are equal to 1 and hence may be ignored in the tensor
product of (\ref{sm}).

The same situation \ one has with another class of integrable models -- low
level WZW models perturbed by current-current perturbation. In the most of
cases they are equivalent to the GN models -- just by fermionization of WZW
models (see, e.g. \cite{DMS}). They are well defined objects, while the GN
model are not always well defined: for example, as we said $C_{n}$ GN model
cannot be defined on the usual Majorana fermions, but level one $C_{n}$ WZW
model with current-current perturbation is well defined and described by the
S-matrix (\ref{sm}) with $l\rightarrow \infty ,m=1$. Here another
interesting identity is valid. Analyzing Boltsmann weights together with the
restriction condition (\ref{res}), one can see the identity (up to a
rescaling of the spectral parameter $u$) between $C_{n}$ level 1 and $%
A_{n-1} $ level 2 RSOS models. This identity, which is absolutely analogous
to the relation $(B_{n})_{1}\symbol{126}(A_{1})_{2}$, was also pointed out
in \cite{G} for affine Lie algebras.

Both of the identities express themselves also in the form of the TBA
diagrams. Looking at the fig 1.b. and at the fig 2.a, in the case $m=1$, one
can see that there remains one ''spurious'' node under the line of massive
nodes on the fig 1.b, and a chain of $n-1$ ''spurious'' nodes under the
massive line on the fig 2.a. They are nothing but the magnonic degrees of
freedom, describing $(A_{1})_{2}$ (TCI) model in the first case, and $%
(A_{n-1})_{2}$ -- in the second. This remark demonstrates how a correct TBA
diagram could help to guess the correct S-matrix.

In \cite{HolS} it was shown, that S-matrix of the form (\ref{gn0}) in the $%
C_{n}$ case sufferes from ''bootstrap violation'': there are double poles
unexplainable by the Coleman-Thun mechanism. As we explained, this confusion
was related with a naive assumption about the S-matrix form (\ref{gn0}),
which was expected to be correct for current-current perturbation of the
level 1 $C_{n}$ WZW model. The correct form of the S-matrix for this
integrable model is $S_{ab}=X_{ab}S_{ab}^{(\infty )}\otimes S_{ab}^{(1)}$
and contains non trivial RSOS tensor factor. As we said, due to the
isomorphism $(C_{n})_{1}\symbol{126}(A_{n-1})_{2}$ \cite{G}, this factor is
physically natural.

\section{Discussion}

We derived TBA equations from the S-matrices (\ref{sm}) for all the infinite
series if Lie algebras $G=A_{n},B_{n},C_{n},D_{n}$. We have shown that with
assumption (\ref{asum}) they give the $Y$-systems with the proper high
temperature behavior reproducing the correct central charge of the cosets.
The assumption (\ref{asum}) was shown to be correct in $A_{n},C_{n}$ cases
for any two fundamental multiplets, and also in $B_{n},D_{n}$ cases for
vector-vector multiplets scattering.

It would be interesting to obtain crossing-unitarising prefactor for other $%
B_{n},D_{n}$ trigonometric RSOS S-matrices with available spectral
decomposition -- spinor-spinor, and vector-spinor ones, in order to be sure
in correctness of the assumption (\ref{asum}) also in these cases.

Derivation of \ the TBA equations presented in this paper is a technical and
quite straightforward procedure. But we think it is not just necessary for
completeness of the CFT - TBA relation picture. As we saw, one gets a
feedback from this derivation procedure important for the form of the
S-matrix itself. An attempt to reduce \ the derived TBA system to a known
and studied $Y$-system, may give a hint for the correct form of S-matrix.
For example, if one would start from the (\ref{gn0}) S-matrix for the $B_{n}$
invariant GN model, he will get the $Y$-system with one missing magnonic
node on the TBA diagram, compared to the correct one (like Fig. 1b). One
immediately realizes what should be added in order to get the correct $Y$%
-system -- tensoring of the S-matrix with RSOS S-matrix of TCI model will
give a desired missing magnonic node. This simple logic may be useful in
consideration of new, less studied integrable models.

\section{Acknowledgements}

I am grateful to Roberto Tateo for many useful discussions and stimulating
interest to this work. I am also thankful to Changrim Ahn and Doron Gepner
for discussions. I thank the Mathematical Department of Durham University
for hospitality, where this work was started. The work was supported by NATO
grant PST.CLG.980424.

\section{Appendix}

\subsection{\protect\bigskip Kernels and mass spectrum}

Here we list the matrices $\widetilde{A}^{G}$ (inverse matrices for $M_{ab}/%
\left[ t_{ab}^{-1}\right] $)

$\mathbf{A}_{n}$%
\begin{equation}
\widetilde{A}_{ab}^{A_{n}}=\frac{\left[ \min (a,b)\right] \left[ n+1-\max
(a,b)\right] }{\left[ n+1\right] }.  \label{a}
\end{equation}

$\mathbf{B}_{n}$%
\begin{eqnarray}
\widetilde{A}_{ab}^{B_{n}} &=&\frac{\left[ \min (a,b)\right] \left( n-\frac{1%
}{2}-\max (a,b)\right) }{\left( n-\frac{1}{2}\right) },\ a,b\leq n-1,
\label{b} \\
\widetilde{A}_{an}^{B_{n}} &=&\widetilde{A}_{na}^{B_{n}}=\frac{\left[ a%
\right] }{2\left( n-\frac{1}{2}\right) },\ a\leq n-1,  \nonumber \\
\widetilde{A}_{nn}^{B_{n}} &=&\frac{\left[ n\right] }{4\left( n-\frac{1}{2}%
\right) (1/2)}.  \nonumber
\end{eqnarray}

$\mathbf{C}_{n}$%
\begin{equation}
\widetilde{A}_{ab}^{C_{n}}=\frac{\left[ \frac{1}{2}\min (a,b)\right] \left(
\frac{1}{2}\left( n+1-\max (a,b)\right) \right) }{\left( \frac{1}{2}\left(
n+1\right) \right) }  \label{c}
\end{equation}

$\mathbf{D}_{n}\mathbf{.}$%
\begin{eqnarray}
\widetilde{A}_{ab}^{D_{n}} &=&\frac{\left[ \min (a,b)\right] \left( n-1-\max
(a,b)\right) }{\left( n-1\right) },\ a,b\leq n-2,  \label{d} \\
\widetilde{A}_{an}^{D_{n}} &=&\widetilde{A}_{an-1}^{D_{n}}=\widetilde{A}%
_{na}^{D_{n}}=\widetilde{A}_{n-1a}^{D_{n}}=\frac{\left[ a\right] }{2\left(
n-1\right) },\ a\leq n-2,  \nonumber \\
\widetilde{A}_{n-1n-1}^{D_{n}} &=&\widetilde{A}_{nn}^{D_{n}}=\frac{\left[ n%
\right] }{4\left( 1\right) \left( n-1\right) },  \nonumber \\
\widetilde{A}_{nn-1}^{D_{n}} &=&\widetilde{A}_{n-1n}^{D_{n}}=\frac{\left[ n-2%
\right] }{4\left( 1\right) \left( n-1\right) },  \nonumber
\end{eqnarray}

We also list here the kernels $K^{G_{n}}$ (their non zero elements), which
are inverse for $A^{G_{n}}$, and differ by a scalar factor from $M_{ab}/%
\left[ t_{ab}^{-1}\right] $:

$\mathbf{A}_{n}$%
\begin{equation}
K_{ab}^{A_{n}}=\delta _{ab}-\frac{1}{2(1)}\left( \delta _{a+1,b}+\delta
_{a-1,b}\right) .  \label{ka}
\end{equation}

$\mathbf{B}_{n}$%
\begin{eqnarray}
K_{ab}^{B_{n}} &=&K_{ab}^{A_{n}},\ a,b\leq n-1,  \label{kb} \\
K_{nn-1}^{B_{n}} &=&K_{n-1n}^{B_{n}}=-\frac{\left( 1/2\right)
}{(1)},  \nonumber
\\
K_{nn}^{B_{n}} &=&f=\frac{2(1/2)^{2}}{(1)}.  \nonumber
\end{eqnarray}

$\mathbf{C}_{n}$%
\begin{eqnarray}
K_{ab}^{C_{n}} &=&fK_{ab}^{A_{n}}\left( \frac{\omega }{2}\right) ,\ a,b\leq
n-1,  \label{kc} \\
K_{nn-1}^{C_{n}} &=&K_{n-1n}^{C_{n}}=-\frac{\left( 1/2\right)
}{(1)},  \nonumber
\\
K_{nn}^{C_{n}} &=&1.  \nonumber
\end{eqnarray}

$\mathbf{D}_{n}$%
\begin{eqnarray}
K_{ab}^{D_{n}} &=&K_{ab}^{A_{n}},\ a,b\leq n-2,  \label{kd} \\
K_{n-1n-2}^{D_{n}} &=&K_{n-1n}^{D_{n}}=K_{nn-1}^{D_{n}}=K_{n-2n-1}^{D_{n}}=-%
\frac{1}{2(1)},  \nonumber \\
K_{n-1n-1}^{D_{n}} &=&K_{nn}^{D_{n}}=1.  \nonumber
\end{eqnarray}

As was mentioned above, mass spectrum vectors $m_{a}$, with components
corresponding to masses of different fundamental representation multiplets
of algebras $G_{n}$, form eigenvectors of matrices $K^{G_{n}}$ listed above
with zero eigenvalues. Here we recall the mass spectra for different algebras

\bigskip

\begin{tabular}{|l|l|l|l|}
\hline
$A_{n}$ & $B_{n}$ & $C_{n}$ & $D_{n}$ \\ \hline
$m_{a}=M\sin \frac{\pi a}{n+1}$ & $m_{a}=2M\sin \frac{\pi a}{2n-1}$ & $M\sin
\frac{\pi a}{2(n+1)}$ & $m_{a}=2M\sin \frac{\pi a}{2n+2}$ \\
& $(a\leq n-1)$ &  & $(a\leq n-2)$ \\
& $m_{n}=M$ &  & $m_{n-1}=m_{n}=M$ \\ \hline
\end{tabular}

\subsection{Solution of (\ref{tmc1}),(\ref{tmc2}).}

First, we express $\rho _{1}^{n-1}$ using the eq. (\ref{c1}) with $j=1$:
\[
\rho _{1}^{n-1}=\sum_{b=1}^{n-1}A_{n-1b}^{A_{n-1}}(\omega /2)\left( \frac{1}{%
2(1)}\sigma ^{b}-\sum_{k=1}^{2l-1}K^{1k}(\omega /2)\widetilde{\rho }%
_{k}^{b}\right).
\]
Hence, using the identity
\[
A_{n-1a}^{A_{n-1}}(\omega /2)=f\frac{A_{n-1a}^{C_{n}}}{A_{nn}^{C_{n}}}=2(1/2)%
\frac{A_{na}^{C_{n}}}{A_{nn}^{C_{n}}},
\]
we can write the last term in (\ref{tmc2}) as
\[
D=\frac{1}{2(1)}\rho _{1}^{n-1}=2(1/2)\sum_{a=1}^{n-1}\frac{A_{na}^{C_{n}}}{%
A_{nn}^{C_{n}}}\left( \frac{1}{2(1)}\sigma
^{a}-\sum_{k=1}^{2l-1}K^{1k}(\omega /2)\widetilde{\rho
}_{k}^{a}\right).
\]
Introducing notations
\[
B^{a}=\frac{(1/2)}{(1)}\left( a_{1/2}^{(l)}\sigma ^{a}-\widetilde{\rho }%
_{1}^{a}\right) ,\ C=\frac{1}{2(1)}\left( a_{1}^{(l)}\sigma ^{n}-\widetilde{%
\rho }_{1}^{n}\right) -D,
\]
the system (\ref{tmc1}),(\ref{tmc2}) may be written as
\begin{eqnarray}
\sum_{b=1}^{n-1}K_{ab}^{C_{n}}x^{b}-\delta _{a,n-1}\frac{(1/2)}{(1)}x^{n}
&=&B^{a} \nonumber\\
-\frac{(1/2)}{(1)}x^{n-1}+x^{n} &=&C. \nonumber
\end{eqnarray}
Using the last equation, one can express $x^{n}$ and substitute it into the
first one getting
\[
\sum_{b=1}^{n-1}\left( K_{ab}^{C_{n}}-\delta _{a,n-1}\delta _{b,n-1}\frac{%
(1/2)^{2}}{(1)^{2}}\right) x^{b}=B^{a}+\delta
_{a,n-1}\frac{(1/2)}{(1)}C.
\]
The matrix in the l.h.s. has as an inverse the restriction of $%
A_{ab}^{C_{n}} $ to the values $a,b\leq n-1$, and we have the solution:
\[
x^{a}=A_{an-1}^{C_{n}}\frac{(1/2)}{(1)}C+\sum_{b=1}^{n-1}A_{ab}^{C_{n}}B^{b}.
\]
Using the obtained $x^{n-1}$ in the second equation, we get
\[
x^{n}=A_{nn}^{C_{n}}C+\frac{(1/2)}{(1)}\sum_{b=1}^{n-1}A_{n-1b}^{C_{n}}B^{b},
\]
where we used the identity
\[
A_{n-1n-1}^{C_{n}}\frac{(1/2)^{2}}{(1)^{2}}+1=A_{nn}^{C_{n}}.
\]
Explicit use of \ the definitions of $B^{a},C,D$ gives the expressions we
used in the main text.

\subsection{Solution of (\ref{tmb1}) - (\ref{tmb3}).}

We solve (\ref{b3}) with respect to $x^{n}$%
\[
x^{n}=\frac{1}{2(1/2)}\left( a_{1/2}^{(l)}\sigma ^{n}-\widetilde{\rho }%
_{1}^{n}\right) -\frac{1}{f}K_{n,n-1}^{B_{n}}x^{n-1},
\]
and substitute it into (\ref{b2}). We get an equation, which together with (%
\ref{b1}), can be written as
\begin{equation}
\sum_{b=1}^{n-1}\left( K_{ab}^{B_{n}}-\delta _{a,n-1}\delta _{b,n-1}\frac{1}{%
f}\left( K_{n,n-1}^{B_{n}}\right) ^{2}\right) x^{b}=B^{a}+\delta _{a,n-1}%
\frac{1}{2(1)}C,  \label{bb}
\end{equation}
where
\begin{eqnarray}
B^{a} &=&\frac{1}{2(1)}\left( a_{1}^{(l)}\sigma ^{a}-\widetilde{\rho }%
_{1}^{a}\right) , \nonumber\\
C &=&a_{1/2}^{(l)}\sigma ^{n}-\widetilde{\rho }_{1}^{n}-\rho
_{1}^{n}. \nonumber
\end{eqnarray}
The inverse matrix for the one in the parenthesis in the l.h.s. of (\ref{bb}%
) coincides with $A_{ab}^{B_{n}}$, and we get
\[
x^{a}=\sum_{b=1}^{n-1}A_{ab}^{B_{n}}B^{b}+A_{an-1}^{B_{n}}C
\]
One can now substitute $x^{n-1}$ in the last form into $x^{n}$ and get
finally the expressions (\ref{xab})(\ref{xnb}), after collection of similar
terms, using the kernel identity $A_{an-1}^{B_{n}}=2(1/2)A_{an}^{B_{n}}$,
valid for $a\leq n-1$.

\end{document}